\documentclass[12pt]{article} 

  \usepackage{graphicx}
  \usepackage{color}
  \newlength{\abstractwidth}
\usepackage[english]{babel}
\usepackage[utf8x]{inputenc}
\usepackage[T1]{fontenc}
\usepackage{braket}
\usepackage{comment}
\usepackage{ dsfont }
\usepackage[title]{appendix}
\usepackage{cite}
\usepackage[a4paper,top=3cm,bottom=2cm,left=3cm,right=3cm,marginparwidth=1.75cm]{geometry}

\usepackage{amsmath}
\usepackage{graphicx}
\usepackage{bbm}
\usepackage{amssymb}
\usepackage[colorinlistoftodos]{todonotes}
\usepackage{caption}
\usepackage{subfigure}
\usepackage{authblk}
\usepackage{float}
\usepackage{inputenc}
\usetikzlibrary{backgrounds,automata}

\def\XXint#1#2#3{{\setbox0=\hbox{$#1{#2#3}{\int}$}
     \vcenter{\hbox{$#2#3$}}\kern-.5\wd0}}

\usepackage{graphicx}
\usepackage{caption}
\usepackage{lipsum}
\usepackage{hyperref}

\usepackage{braket}

\begin{document}

\begin{titlepage}
 % \rightline{}
  \bigskip

  \bigskip\bigskip

  \bigskip

\begin{center}
%\centerline
{\Large \bf{}}
 \bigskip
%\centerline
{\Large \bf {Replica wormhole and information retrieval in the}} 
~\\
~\\
{\Large \bf { SYK model coupled to Majorana chains}} 
\bigskip
\bigskip
   \bigskip
\bigskip
\end{center}

  \begin{center}

 \bf {Yiming Chen$^{1}$, Xiao-Liang Qi$^{2-3}$ and Pengfei Zhang$^{4-5}$}
  \bigskip \rm
\bigskip

$^{1}$Jadwin Hall, Princeton University,  Princeton, NJ 08540, USA\\
\bigskip
$^{2}$Stanford Institute for Theoretical Physics, Stanford University, Stanford, CA 94305, USA \\
\bigskip
$^{3}$Department of Physics,, Stanford University, Stanford, CA 94305, USA \\
\bigskip
$^{4}$Walter Burke Institute for Theoretical Physics, California Institute of Technology, Pasadena, CA 91125, USA \\ 
\bigskip
$^{5}$Institute for Quantum Information and Matter, California Institute of Technology, Pasadena, CA 91125, USA\\ 
% \vspace{2cm}
  \end{center}

 \bigskip\bigskip
  \begin{abstract}
Motivated by recent studies of the information paradox in (1+1)-D anti-de Sitter spacetime with a bath described by a (1+1)-D conformal field theory, we study the dynamics of second R\'{e}nyi entropy of the Sachdev-Ye-Kitaev (SYK) model ($\chi$) coupled to a Majorana chain bath ($\psi$). The system is prepared in the thermofield double (TFD) state and then evolved by $H_L+H_R$. For small system-bath coupling, we find that the second R\'{e}nyi entropy $S^{(2)}_{\chi_L, \chi_R}$ of the SYK model undergoes a first order transition during the evolution. In the sense of holographic duality, the long-time solution corresponds to a ``replica wormhole''. The transition time corresponds to the Page time of a black hole coupled to a thermal bath. We further study the information scrambling and retrieval by introducing a classical control bit, which controls whether or not we add a perturbation in the SYK system. The mutual information between the bath and the control bit shows a positive jump at the Page time, indicating that the entanglement wedge of the bath includes an island in the holographic bulk.
 \end{abstract}
\bigskip \bigskip \bigskip

  \end{titlepage}

   % \starttext \baselineskip=17.63pt \setcounter{footnote}{0}
   \tableofcontents
\newpage
 % \sc

\section{Introduction}

The black hole information paradoxes refer to various kinds of obstruction in combining black hole gravitational physics and quantum mechanics. As a well known example, Hawking’s calculation \cite{hawking1975particle} for the entropy of the radiation from a pure state evaporating black hole leads to a monotonically growing result, which is inconsistent with an unitary evolution where the late time entropy is expected to follow a Page curve \cite{page1993average}. 

To compute entropy in holographic systems, a powerful tool is provided by the Ryu-Takayanagi (RT) formula \cite{Ryu:2006bv,Ryu:2006ef,Lewkowycz:2013nqa}. The formula was first proposed for stationary asymptotic anti-de-Sitter (AdS) spacetime, and has been subsequently generalized to time-dependent cases, known as the Hubeny-Rangamani-Ryu-Takayanagi (HRRT) formula \cite{Hubeny:2007xt}. After taking the contributions from the bulk quantum fields into account \cite{Faulkner:2013ana,Engelhardt:2014gca}, the general proposal states that the von Neumann entropy of a boundary region $A$ is determined by finding all extremums of the generalized entropy:
\begin{align}\label{RT}
S_{vN} (A) = \textrm{min} \left[ \textrm{ext}_{\gamma_A} \left( \frac{\textrm{Area}(\gamma_{A})}{4G_N} + S_{bulk}\right)\right],
\end{align}
among all possible bulk surfaces $\gamma_A$ that are homologous to $A$ and then look for the minimal one. In the formula, $S_{bulk}$ is the entropy of the bulk quantum fields in the region bounded by the quantum extremal surface $\gamma_A$ and the boundary. 

Recently, new insights on the information paradox have been brought by discovering a new quantum extremal surface for an evaporating AdS black hole \cite{penington2019entanglement,almheiri2019entropy}. As emphasized in \cite{almheiri2019page,almheiri2019islands}, when applying  eq. \eqref{RT} to calculate the entropy of the radiation, one must consider possible solutions involving entanglement "islands" in the bulk. In most of these papers, the set-up is to allow the black hole evaporating into a bath by gluing the AdS boundary with an auxiliary spacetime with no gravitational degree of freedom. Before the Page time, the quantum extremal surface of the radiation in the bath is trivial, and the entropy of the radiation agrees with Hawking's field theory calculation. In contrast, after the Page time the quantum extremal surface becomes nontrivial and bounds an isolated island in the bulk. The first order transition between these two solutions gives the Page curve. In \cite{almheiri2019replica,penington2019replica}, these new solutions are explained as coming from the replica wormhole solutions in the gravitational path integral derivation of the entropy, and the transition becomes a cross-over if one sums over all geometries in the model of \cite{penington2019replica}. See also recent discussions on the information paradox, islands and replica wormholes in \cite{Akers:2019wxj,Fu:2019oyc,Akers:2019nfi,Rozali:2019day,Chen:2019uhq,Bousso:2019ykv,Almheiri:2019psy,Chen:2019iro,Marolf:2020xie}. Since the quantum extremal surface of the radiation bounds an island in the bulk, the entanglement wedge reconstruction proposal \cite{dong2016reconstruction} implies that the information inside the island, which covers part of the black hole interior, should become accessible to the bath after the Page time \cite{hayden2007black}. Concrete ways to recover the information inside the island have been proposed via the Petz map \cite{penington2019replica} or the modular flow \cite{Chen:2019iro}. 

In this paper, we study the black hole evaporation problem by considering a Sachdev-Ye-Kitaev (SYK) model coupled with a $(1+1)$-dimensional free fermion bath. The SYK model \cite{maldacena2016conformal,kitaev2018soft} is a $(0+1)$-dimensional strongly correlated fermion model with emergent nearly conformal dynamics at low temperature. The low energy dynamics of the SYK model has a holographic dual theory which is the Jackiw-Teitelboim gravity in AdS$_2$ \cite{maldacena2016conformal,kitaev2018soft}. %which is a UV-complete quantum mechanical model for random interacting Majorana fermions. Motivated by above developments on the gravity side, we would like to study the counterpart of above transitions of the entanglement entropy in the context of the SYK model coupled to a bath. 
Previously, the physics of a SYK model coupled to a large SYK bath has been studied in \cite{chen2017tunable,zhang2019evaporation,almheiri2019universal}, while the entropy dynamics of the SYK model have been studied in \cite{gu2017spread,penington2019replica} using coupled SYK model with equal number of modes. A new saddle point solution after the Page time, which corresponds to the replica wormhole, has been discovered in the micro-canonical ensemble in \cite{penington2019replica}. In this work, we instead model the bath by free Majorana chains. Having the simple free dynamics in the bath is helpful in simplifying the problem, and is also closer to the setups with AdS black holes coupled with non-gravitational flat space bath.

For simplicity we consider a thermofield double state of the SYK model coupled to free fermion bath, which then contains two SYK models (left and right) and two baths (also left and right). Denoting the Hamiltonian of each side as $H_L$ and $H_R$ respectively, the thermofield double state is invariant under the time evolution of $H_L-H_R$. We consider the time evolution by $H_L+H_R$ which changes the state and leads to increase of the entanglement between the SYK sites and the baths. This corresponds to a setup with a two-sided eternal black hole in equilibrium with two flat baths, as has been discussed in Ref. \cite{almheiri2019islands}. The set-up is discussed in detail in \textbf{section \ref{sec:setup}}. Using the Schwinger-Dyson equations in the large $N$ limit, we study the second R\'{e}nyi entropy of the union of the two baths in the time-evolved thermofield double state. Equivalently, this can be expressed in terms of the correlation function of two twist operators. As explained in \textbf{section \ref{sec:numerics}}, for small system-bath coupling, numerically we find that the second R\'{e}nyi entropy $S^{(2)}_{\chi_L, \chi_R}$ shows a first-order transition. The short-time saddle can be studied using perturbation theory (\textbf{section \ref{analyticshort}}) and the long-time solution can be explained by twist operator factorization (\textbf{section \ref{sec:analyticlong}}). By introducing a classical control bit, we could ask whether the information thrown into the SYK system can be extracted from the bath by looking at the mutual information between the bath and the control bit. This is discussed in \textbf{section \ref{sec:retrieval}}. We find that the mutual information has a jump at the Page time, which signals that there is an "island" outside the horizon in the gravity picture. Finally, in \textbf{section \ref{sec:conclusion}}, we conclude our work and discuss some open questions. %which is hard to realize by SYK models.

\section{Set-up}
\label{sec:setup}
    
    The SYK model \cite{kitaev2015simple,maldacena2016remarks} describes $N$ Majorana fermion modes $\chi_i$ labeled by $i=1,2,...,N$ with random interaction. We consider coupling each mode $\chi_i$ to an individual (1+1)-d free Majorana chain $\psi_i(x)$ with constant hopping $\Lambda/2$ and periodic boundary condition. These Majorana chains serve as a thermal bath for the SYK system. In most of the following discussion, we will refer to $\chi$ as the system (or the black hole in the gravity analogy), while $\psi$ as the bath. The Hamiltonian of the coupled system is the following:
        \begin{align}\label{model1H}
        H=H_{\chi}+H_{\psi}+H_{\text{int}}=\sum_{i,j,k,l}\frac{J_{ijkl}}{4!}\chi_i\chi_j\chi_k\chi_l+i \sum_{x,i}\frac{\Lambda}{2}\psi_i(x)\psi_i(x+1)+H_{\text{int}},
    \end{align}
 where $x\in \mathds{Z}$ labels different sites in the Majorana chain, and the interaction term $H_{\text{int}}$ will be specified later. We choose the convention that $\{\chi_i,\chi_j\}=\delta_{ij}$ and $\{\psi_i(x),\psi_j(y)\}=\delta_{ij}\delta_{xy}$. $J_{ijkl}$ are Gaussian random variables with the mean and variance:
        \begin{align}
        \overline{J_{ijkl}}=0,\ \ \ \ \ \ \  \overline{(J_{ijkl})^2}=\frac{3!J^2}{N^3}.
        \end{align}
        One can perform a Fourier transform on the Majorana chain: 
            \begin{align}
        \psi_j(x)=\frac{1}{\sqrt{N_L}}\sum_k e^{ik x}\psi_j(k),
    \end{align}
    where $N_L$ is the total number of sites. This gives
    \begin{align}
        \sum_xi \frac{\Lambda}{2}\psi_i(x)\psi_i(x+1)=\sum_{0\leq k \leq \frac{\pi}{a}}\Lambda\sin (k a) \psi_i^\dagger(k)\psi_i(k)\equiv\sum_{0 \leq k \leq \frac{\pi}{a}}\epsilon_k \psi_i^\dagger(k)\psi_i(k).
    \end{align}
    the summation is over a half of the first Brillouin zone. Here $a$ is the lattice spacing of the Majorana chain. In the low energy limit, the bath contains a left-moving Majorana mode with $k\sim \pi/a$ and a right-moving Majorana mode with $k\sim 0$. For simplicity, we choose $a=1/\Lambda$. As a result, in the continuum limit $\Lambda \rightarrow \infty$, we have $\epsilon_k\sim \pm k$ near the gapless points. The central charge $c$ of the conformal field theory for the bath in the continuum limit, which contains $N$ copies of the Majorana chains, is then $N/2$.

  In this work, we choose the interaction term $H_{\text{int}}$ as a hopping of the SYK fermion to the center site $x=0$ of the Majorana chain:
\begin{align}
  H_{\text{int}}=\sum_{i}iV\sqrt{\Lambda}\chi_i\psi_i(0) .
\end{align}
Here we have introduced a factor of $\sqrt{\Lambda}$ since $\sqrt{\Lambda}\psi_i(x)\equiv \eta_i(x)$ corresponds to the continuous fermion operator in the limit $\Lambda \rightarrow \infty$, with the Dirac $\delta$-function anti-commutator $\left\{\eta_i(x),\eta_j(y)\right\}=\delta_{ij}\delta(x-y)$. 

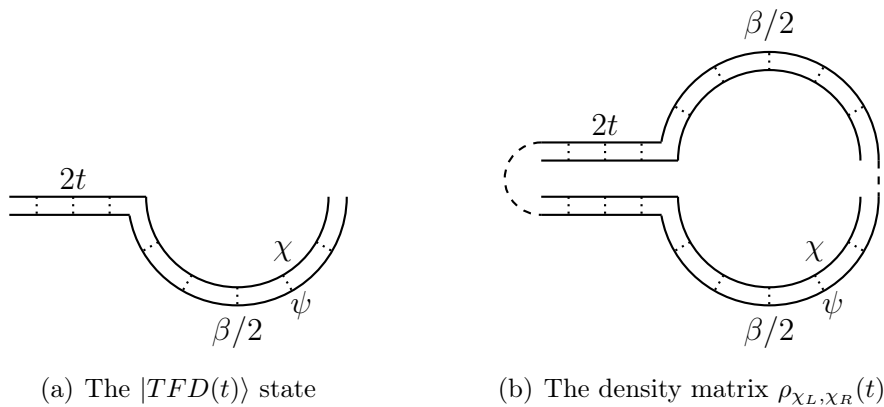
\begin{figure}[t!]  
\centering  

\subfigure[The $\ket{TFD(t)}$ state]  
{  
\begin{tikzpicture}[thick,scale = 1.2]
  \draw (-1,-0.2) -- (-2.5,-0.2);
   \draw (1,-0.2) arc(0:-180:1 and 1);
   \draw (1.2,-0.2) arc(0:-170:1.2 and 1.2);

    \draw (-1.18,-0.4) -- (-2.5,-0.4);

       \draw[dotted] (0.866,-0.7) -- (1.039,-0.8);

        \draw[dotted] (0.5,-1.066) -- (0.6,-1.239);

              \draw[dotted] (0,-1.2) -- (0,-1.4);

        \draw[dotted] (-0.5,-1.066) -- (-0.6,-1.239);

           \draw[dotted] (-0.866,-0.7) -- (-1.039,-0.8);

             \draw[dotted] (-1.4,-0.2) -- (-1.4,-0.4);

             \draw[dotted] (-1.8,-0.2) -- (-1.8,-0.4);

             \draw[dotted] (-2.2,-0.2) -- (-2.2,-0.4);
      \draw (0.5,-0.8) node{$\chi$};
       \draw (0.7,-1.4) node{$\psi$};
       \draw (0,-1.7) node{$\beta/2$};
       \draw (-1.8,0) node{$2t$};
\end{tikzpicture}

}  
\hspace{0.1\textwidth}
\subfigure[The density matrix $\rho_{\chi_L, \chi_R} (t)$]  
{  

\begin{tikzpicture}[thick,scale = 1.2]
 \draw (1,0.2) arc(0:180:1 and 1);
 \draw (1,-0.2) arc(0:-180:1 and 1);
  \draw (-1,-0.2) -- (-2.5,-0.2);
   \draw (-1,0.2) -- (-2.5,0.2);

   \draw (1.2,0.2) arc(0:170:1.2 and 1.2);
   \draw (1.2,-0.2) arc(0:-170:1.2 and 1.2);
   \draw[dashed] (1.2,0.2) --(1.2,-0.2) ;
    \draw (-1.18,-0.4) -- (-2.5,-0.4);
     \draw (-1.18,0.4) -- (-2.5,0.4);
     \draw[dashed] (-2.5,0.4) arc (90:270:0.4 and 0.4);
     \draw[dotted] (0.866,0.7) -- (1.039,0.8);
       \draw[dotted] (0.866,-0.7) -- (1.039,-0.8);
      \draw[dotted] (0.5,1.066) -- (0.6,1.239);
        \draw[dotted] (0.5,-1.066) -- (0.6,-1.239);
           \draw[dotted] (0,1.2) -- (0,1.4);
              \draw[dotted] (0,-1.2) -- (0,-1.4);
          \draw[dotted] (-0.5,1.066) -- (-0.6,1.239);
        \draw[dotted] (-0.5,-1.066) -- (-0.6,-1.239);
         \draw[dotted] (-0.866,0.7) -- (-1.039,0.8);
           \draw[dotted] (-0.866,-0.7) -- (-1.039,-0.8);
           \draw[dotted] (-1.4,0.2) -- (-1.4,0.4);
             \draw[dotted] (-1.4,-0.2) -- (-1.4,-0.4);
                     \draw[dotted] (-1.8,0.2) -- (-1.8,0.4);
             \draw[dotted] (-1.8,-0.2) -- (-1.8,-0.4);
              \draw[dotted] (-2.2,0.2) -- (-2.2,0.4);
             \draw[dotted] (-2.2,-0.2) -- (-2.2,-0.4);
      \draw (0.5,-0.8) node{$\chi$};
       \draw (0.7,-1.4) node{$\psi$};
       \draw (0,-1.7) node{$\beta/2$};
       \draw (0,1.7) node{$\beta/2$};
       \draw (-1.8,0.6) node{$2t$};
\end{tikzpicture}

}

\caption{The graphical representations of (a) the $\ket{TFD(t)}$ state and (b) the reduced density operator $\rho_{\chi_L, \chi_R} (t)$.}
\label{fig:fig1}
\end{figure}

The set up of the problem is as follows. 
We first introduce two copies of the coupled system -- Left: $\chi_L$, $\psi_L$ and Right: $\chi_R$, $\psi_R$, and prepare them in a thermofield doubled (TFD) state \cite{israel1976thermo} with inverse temperature $\beta$. When we only look at the left or right system, it is in a thermal density matrix with Hamiltonian given by (\ref{model1H}), while the whole system is in a pure state. 
The definition of the thermofield double state in this model is not unique. Without losing generality, we make the following explicit choice. We begin by constructing the state $\ket{I}_{\chi_L,\chi_R}$, which satisfies
\begin{equation}
    \left( \chi_{L,j} + i \chi_{R,j} \right)\ket{I}_{\chi_L, \chi_R} = 0, \quad j = 1,2,...,N.
\end{equation}
$\ket{I}_{\chi_L,\chi_R}$ is a maximally entangled state between the $\chi_L$ system and $\chi_R$ system. Similarly, we construct a maximally entangled state $\ket{I}_{\psi_L, \psi_R}$ between the $\psi_L$ system and $\psi_R$ system with spatial locality, which satisfies
\begin{equation}
     \left( \psi_{L,j}(x) + i \psi_{R,j}(x) \right)\ket{I}_{\psi_L, \psi_R} = 0, \quad j = 1,2,...,N;\ \ \ \ x\in \mathds{Z}.
\end{equation}
The thermofield double state is then given by
        \begin{align}
            \ket{TFD}\equiv \frac{e^{-\beta (H_L + H_R ) /4}}{\sqrt{Z(\beta)}} |I\rangle_{\chi_L,\chi_R}|I\rangle_{\psi_L,\psi_R},
        \end{align}
        where $H_L$ and $H_R$ are the Hamiltonian \eqref{model1H} defined on the left and right system. After we have the state $\ket{TFD}$, we evolve the system in time using $H_L+H_R$. One important property of the thermofield double state is that it is annihlated by $H_L - H_R$, and thus we can pull all the evolution on the right system onto the left system, and write the time-evolved TFD state as:
        \begin{align}
            |TFD(t)\rangle=\frac{e^{-i2 H_L t}e^{-\beta H_L/2}}{\sqrt{Z(\beta)}}|I\rangle_{\chi_L,\chi_R}|I\rangle_{\psi_L,\psi_R}.
        \end{align}
        The $\ket{TFD(t)}$ state can be represented graphically as in fig. \ref{fig:fig1} (a). The inner/outer line represents the $\chi/\psi$ system. We have suppressed the extra spatial dimension for $\psi$. The half circle with length $\beta/2$ corresponds to the Euclidean preparation for the $\ket{TFD}$ state, followed by a real time evolution of $2t$ represented by the horizontal lines. The dotted lines between $\chi$ and $\psi$ denote the interaction in the system.

        In this paper, we will focus on calculating the 
       second R\'{e}nyi entropy of subsystem $\chi_L \cup \chi_R$. This is in analogy of calculating the entropy for the black holes in \cite{almheiri2019islands}. Due to the inherent unitarity here, this is the same as calculating the second R\'{e}nyi entropy of the baths, since we started from a pure state. This is different from the gravity story, where \emph{a priori} one cannot assume unitarity when trying to address the information paradox.  
       
       The reduced density matrix of subsystem $\chi_L \cup \chi_R$ is given by:
        \begin{align}
            \rho_{\chi_L,\chi_R}(t)=\text{tr}_{\psi_L,\psi_R}|TFD(t)\rangle\langle TFD(t)|.
        \end{align}
       We draw the graphical representation of the density matrix in fig. \ref{fig:fig1}(b). We take two copies of the state in fig. \ref{fig:fig1}(a), one for the ket and one for the bra, and then trace out the $\psi_L \cup \psi_R$ system (denoted by the dashed lines in Fig. \ref{fig:fig1}(b)).

        The $n$-th R\'{e}nyi entropy of a density matrix $\rho$ is defined as
        \begin{equation}
            S^{(n)} \equiv \frac{1}{1-n} \log \textrm{tr} \rho^n.
        \end{equation}
        Specifically, we are interested in the second R\'{e}nyi entropy of the density matrix $ \rho_{\chi_L,\chi_R}(t)$, given by
        \begin{align}\label{Renyi}
            \exp\left(-S^{(2)}_{\chi_L, \chi_R}(t)\right)=\text{tr}\left(\rho_{\chi_L,\chi_R}(t)^2\right).
        \end{align}

        Equivalently, we could express the right hand side as the expectation of twist operators on two copies of the coupled system on state $|TFD(t)\rangle\otimes|TFD(t)\rangle$:
        \begin{align}
        \text{tr}\left(\rho_{\chi_L,\chi_R}(t)^2\right)= \left<T_LT_R\right>.
        \end{align}
        Here the twist operator $T_{L/R}$ operates on the two copies of the $\chi_{L/R}$ system by swapping their states:
       \begin{align}
        T_a|\Psi_1\rangle_{\chi_a}\otimes|\Psi_2\rangle_{\chi_a}=|\Psi_2\rangle_{\chi_a}\otimes|\Psi_1\rangle_{\chi_a},\quad a = L/R.
       \end{align}

\begin{figure}  
\centering  

\subfigure[]  
{  
\begin{tikzpicture}[thick,scale = 1]
 \draw (1,0.2) arc(0:180:1 and 1);
 \draw (1,-0.2) arc(0:-180:1 and 1);
  \draw (-1,-0.2) -- (-2.5,-0.2);
   \draw (-1,0.2) -- (-2.5,0.2);

   \draw (1.2,0.2) arc(0:170:1.2 and 1.2);
   \draw (1.2,-0.2) arc(0:-170:1.2 and 1.2);
   \draw[dashed] (1.2,0.2) --(1.2,-0.2) ;
    \draw (-1.18,-0.4) -- (-2.5,-0.4);
     \draw (-1.18,0.4) -- (-2.5,0.4);
     \draw[dashed] (-2.5,0.4) arc (90:270:0.4 and 0.4);
     \draw[dotted] (0.866,0.7) -- (1.039,0.8);
       \draw[dotted] (0.866,-0.7) -- (1.039,-0.8);
      \draw[dotted] (0.5,1.066) -- (0.6,1.239);
        \draw[dotted] (0.5,-1.066) -- (0.6,-1.239);
           \draw[dotted] (0,1.2) -- (0,1.4);
              \draw[dotted] (0,-1.2) -- (0,-1.4);
          \draw[dotted] (-0.5,1.066) -- (-0.6,1.239);
        \draw[dotted] (-0.5,-1.066) -- (-0.6,-1.239);
         \draw[dotted] (-0.866,0.7) -- (-1.039,0.8);
           \draw[dotted] (-0.866,-0.7) -- (-1.039,-0.8);
           \draw[dotted] (-1.4,0.2) -- (-1.4,0.4);
             \draw[dotted] (-1.4,-0.2) -- (-1.4,-0.4);
                     \draw[dotted] (-1.8,0.2) -- (-1.8,0.4);
             \draw[dotted] (-1.8,-0.2) -- (-1.8,-0.4);
              \draw[dotted] (-2.2,0.2) -- (-2.2,0.4);
             \draw[dotted] (-2.2,-0.2) -- (-2.2,-0.4);
      \draw (0.5,-0.8) node{$\chi$};
       \draw (0.7,-1.4) node{$\psi$};
 
 \begin{scope}[shift={(0,3)}]
 \draw (1,0.2) arc(0:180:1 and 1);
 \draw (1,-0.2) arc(0:-180:1 and 1);
  \draw (-1,-0.2) -- (-2.5,-0.2);
   \draw (-1,0.2) -- (-2.5,0.2);

   \draw (1.2,0.2) arc(0:170:1.2 and 1.2);
   \draw (1.2,-0.2) arc(0:-170:1.2 and 1.2);
   \draw[dashed] (1.2,0.2) --(1.2,-0.2) ;
    \draw (-1.18,-0.4) -- (-2.5,-0.4);
     \draw (-1.18,0.4) -- (-2.5,0.4);
     \draw[dashed] (-2.5,0.4) arc (90:270:0.4 and 0.4);
     \draw[dotted] (0.866,0.7) -- (1.039,0.8);
       \draw[dotted] (0.866,-0.7) -- (1.039,-0.8);
      \draw[dotted] (0.5,1.066) -- (0.6,1.239);
        \draw[dotted] (0.5,-1.066) -- (0.6,-1.239);
           \draw[dotted] (0,1.2) -- (0,1.4);
              \draw[dotted] (0,-1.2) -- (0,-1.4);
          \draw[dotted] (-0.5,1.066) -- (-0.6,1.239);
        \draw[dotted] (-0.5,-1.066) -- (-0.6,-1.239);
         \draw[dotted] (-0.866,0.7) -- (-1.039,0.8);
           \draw[dotted] (-0.866,-0.7) -- (-1.039,-0.8);
           \draw[dotted] (-1.4,0.2) -- (-1.4,0.4);
             \draw[dotted] (-1.4,-0.2) -- (-1.4,-0.4);
                     \draw[dotted] (-1.8,0.2) -- (-1.8,0.4);
             \draw[dotted] (-1.8,-0.2) -- (-1.8,-0.4);
              \draw[dotted] (-2.2,0.2) -- (-2.2,0.4);
             \draw[dotted] (-2.2,-0.2) -- (-2.2,-0.4);
      \draw (0.5,-0.8) node{$\chi$};
       \draw (0.7,-1.4) node{$\psi$};
       
\end{scope}
\draw[dashed] (-2.5,3+0.2) arc (90:270:1 and 1.7);
\draw[dashed] (-2.5,3-0.2) arc (90:270:0.5 and 1.3);
\draw[dashed] (1,3+0.2) arc (90:-90:1 and 1.7);
\draw[dashed] (1,3-0.2) arc (90:-90:0.5 and 1.3);

 \draw (-2.5,3+0.6-0.03) node[circle,fill,scale = 0.1,above]{A};
 \draw[->] (-2.5, 3+0.6) -- (-1.5, 3+0.6) ;
 \draw (-2.3,3+0.9) node{$s=0$} ;
 
 \draw (-2.5,0.6-0.03) node[circle,fill,scale = 0.1,above]{A};
 \draw[->] (-2.5, 0.6) -- (-1.5, 0.6) ;
 \draw (-2.1,0.9) node{$s=\beta + 4t$} ;
 
\end{tikzpicture}

}  
\hspace{0.1\textwidth}
\subfigure[]  
{  

\begin{tikzpicture}[thick,scale = 0.8]
\draw (0.1,0.3+0.1) arc (0:-180:0.1 and 0.1);
\draw (0.3+0.1,0.1) arc (90:270:0.1 and 0.1);
\draw (-0.3-0.1,-0.1) arc (-90:90:0.1 and 0.1);
\draw (0.1,-0.3-0.1) arc (0:180:0.1 and 0.1);

\draw  (0.1,0.3+0.1)  --  (0.1,0.3+0.1+1.2);
\draw  (-0.1,0.3+0.1)  --  (-0.1,0.3+0.1+1.2);  
\draw (0.3+0.1,0.1) --(0.3+0.1+1.2,0.1) ;
\draw  (0.3+0.1,-0.1) --(0.3+0.1+1.2,-0.1) ;
\draw (-0.3-0.1,0.1) --(-0.3-0.1-1.2,0.1) ;
\draw  (-0.3-0.1,-0.1) --(-0.3-0.1-1.2,-0.1) ;
\draw  (0.1,-0.3-0.1)  --  (0.1,-0.3-0.1-1.2);
\draw  (-0.1,-0.3-0.1)  --  (-0.1,-0.3-0.1-1.2);

\draw  (0.1,0.3+0.1+1.2) arc (-90+5.74:270-5.74:1 and 1);
\draw  (0.3+0.1+1.2,0.1)  arc (180-5.74:-180+5.74:1 and 1);
\draw  (-0.3-0.1-1.2,0.1)  arc (5.74:360-5.74:1 and 1);
\draw  (-0.1,-0.3-0.1-1.2) arc (90+5.74:450-5.74:1 and 1);

\draw[dotted]  (0.1,0.3+0.1+0.3) arc (90:0:0.6 and 0.6 );
\draw[dotted]   (0.3+0.1+0.3,-0.1) arc (0:-90:0.6 and 0.6 );
\draw[dotted]   (-0.3-0.1-0.3,0.1)  arc (180:90:0.6 and 0.6 );
\draw[dotted]   (-0.3-0.1-0.3,-0.1)  arc (180:270:0.6 and 0.6 );

\draw[dotted]  (0.1,0.3+0.1+0.6) arc (90:0:0.9 and 0.9 );
\draw[dotted]   (0.3+0.1+0.6,-0.1) arc (0:-90:0.9 and 0.9 );
\draw[dotted]   (-0.3-0.1-0.6,0.1)  arc (180:90:0.9 and 0.9 );
\draw[dotted]   (-0.3-0.1-0.6,-0.1)  arc (180:270:0.9 and 0.9 );

\draw[dotted]  (0.1,0.3+0.1+0.9) arc (90:0:1.2 and 1.2 );
\draw[dotted]   (0.3+0.1+0.9,-0.1) arc (0:-90:1.2 and 1.2 );
\draw[dotted]   (-0.3-0.1-0.9,0.1)  arc (180:90:1.2 and 1.2 );
\draw[dotted]   (-0.3-0.1-0.9,-0.1)  arc (180:270:1.2 and 1.2 );

\draw[dotted]  (0.5, 1.73) arc (90-16.12:16.12:1.8 and 1.8 );
\draw[dotted]  (-0.5, 1.73) arc (90+16.12:180-16.12:1.8 and 1.8 );
\draw[dotted]  (0.5, -1.73) arc (-90+16.12:-16.12:1.8 and 1.8 );
\draw[dotted]  (-0.5, -1.73) arc (270-16.12:180+16.12:1.8 and 1.8 );

\draw[dotted]  (1, 2.25) arc (90-23.96:23.96:2.46 and 2.46 );
\draw[dotted]  (-1, 2.25) arc (90+23.96:180-23.96:2.46 and 2.46 );
\draw[dotted]  (1,-2.25) arc (-90+23.96:-23.96:2.46 and 2.46 );
\draw[dotted]  (-1,-2.25) arc (270-23.96:180+23.96:2.46 and 2.46 );

\draw[dotted]  (0.6, 3.4) arc (90-10:10:3.45 and 3.45);
\draw[dotted]  (-0.6, 3.4) arc (90+10:180-10:3.45 and 3.45);
\draw[dotted]  (0.6,-3.4) arc (-90+10:-10:3.45 and 3.45);
\draw[dotted]  (-0.6, -3.4) arc (270-10:180+10:3.45 and 3.45);

\draw (0,3.2) node{$\chi$}; 
\draw (0,-3.2) node{$\chi$}; 
\draw (3.2,0) node{$\psi$}; 
\draw (-3.2,0) node{$\psi$}; 
\end{tikzpicture}

}

\caption{Two equivalent illustrations of the contour $C$ of path integral for computing the second R\'{e}nyi entropy.}
\label{fig:fig2}
\end{figure}
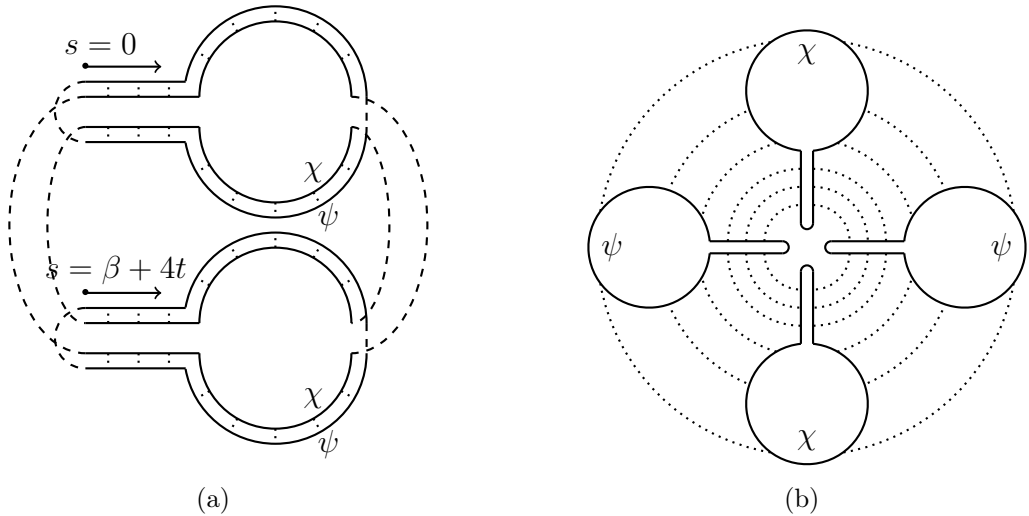

We can formulate the calculation of the second R\'{e}nyi entropy in terms of a path-integral over a replicated contour $\mathcal{C}$ with twisted boundary conditions. The contour $\mathcal{C}$ is shown in fig. \ref{fig:fig2}(a), where we take two copies of the density matrix in fig. \ref{fig:fig1}(b), and join the open ends of the $\chi$ systems in a twisted way (denoted by the dashed lines). In this figure, we've also marked how we parametrize the contour using a real parameter $s$: $s\in [0,\beta+4t)$ covers the upper part of the contour in the clockwise direction, while $s\in (\beta + 4t, 2\beta + 8t]$ covers the lower part of the contour, also in the clockwise direction. The parameterization will be needed later in the presentation of our numerical results. Another equivalent way to picture the contour $\mathcal{C}$ is in fig. \ref{fig:fig2}(b), where it makes clear that the replica contour has the topology of four circles.

The path integral has the following form
\begin{equation}\label{singlerealize}
       e^{-S^{(2)}_{\chi_L,\chi_R}}= \frac{1}{Z^2}\int \mathcal{D}\chi(s)\mathcal{D}\psi (s,x) \exp(-S_\mathcal{C}[\chi,\psi]),
\end{equation}
\begin{equation}
\begin{aligned}
    S_\mathcal{C}[\chi,\psi] =  &\int_{\mathcal{C}}   ds\, \left( \sum_{i} \frac{1}{2}\chi_i \partial_s \chi_i +  \sum_{i,x} \frac{1}{2}\psi_i(x) \partial_s \psi_i(x) \right. \\
    & \left. + f(s) \left(\sum_{i,j,k,l}\frac{J_{ijkl}}{4!}\chi_i\chi_j\chi_k\chi_l+i \sum_{x,i}\frac{\Lambda}{2}\psi_i(x)\psi_i(x+1) +\sum_{i}iV\sqrt{\Lambda}\chi_i\psi_i(0) \right)\right).
\end{aligned}
\end{equation}
Because we are using a single real parameter $s$ to label the contour, we need to introduce an extra factor $f(s)$ to account for whether we are doing imaginary time evolution ($f(s) =1$), forward real time evolution ($f(s) =i$) or backward real time evolution ($f(s) =-i$) \footnote{More explicitly, by the parametrization in fig. \ref{fig:fig2}(a), $f(s)$ is defined as
\begin{equation}
    f(s) \equiv \left\{
    \begin{aligned}
    -i, \quad & s \in (0,2t) \cup (\beta + 4t , \beta + 6t), \\
    i, \quad & s \in (\beta  +2t,  \beta + 4t) \cup (2\beta  +6t,  2\beta + 8t) ,\\
    1, \quad & s \in (2t,  \beta + 2t) \cup (\beta  +6t,  2\beta + 6t) .\\
    \end{aligned}
    \right.
\end{equation}
}.
The expression (\ref{singlerealize}) applies for a single realization of the SYK Hamiltonian. However, in order to apply the standard large $N$ technique of the SYK model, one has to average over the disorder coupling, and approximate $S^{(2)}_{\chi_L, \chi_R}(t)$ by the disorder-averaged value: \footnote{There have been many recent discussions in gravity about the role of disorder/ensemble average and its relation with replica wormholes, see \cite{Saad:2018bqo,penington2019replica,Marolf:2020xie} for examples. The "replica wormhole" solution that we will discuss below does not rely on the disorder average, since different replicas are directly coupled together as in fig. \ref{fig:fig2}(b).}
\begin{equation}
  S^{(2)}_{\chi_L, \chi_R}(t) \approx   \overline{\log(\text{tr}\rho_{\chi_L,\chi_R}^2)} \approx \log(\overline{\text{tr}\rho_{\chi_L,\chi_R}^2}).
\end{equation}
The second approximation comes from the assumption that the dominant saddle point remains replica diagonal. It should be noted that there are two different kinds of replica discussed here. In computing $\overline{\left({\rm tr}(\rho^2)\right)^k}$, there are $2k$ replica labeled by $s=1,2,~\alpha=1,2,...,k$. We assume the dominant saddle point is diagonal in $\alpha$, so that $\overline{\left({\rm tr}(\rho^2)\right)^k}\simeq \overline{\left({\rm tr}(\rho^2)\right)}^k$. In general, the solution is off-diagonal in $s$ which labels the two replica we discussed above in computing $\overline{{\rm tr}(\rho^2)}$.

After the standard procedure of introducing the bilocal $\tilde{G},\tilde{\Sigma}$ fields,  and integrating out the fermion fields $\chi$ and $\psi(x)$, one arrives at
\begin{equation}
               e^{-S^{(2)}_{\chi_L,\chi_R}}= \frac{1}{Z^2}\int \mathcal{D}\tilde{\Sigma} \mathcal{D}\tilde{G} \exp(-S_\mathcal{C}[\tilde{\Sigma},\tilde{G}]),
\end{equation}
\begin{equation}
\begin{aligned}\label{Seff}
           S_\mathcal{C}[\tilde{\Sigma},\tilde{G}]&= - \frac{N}{2} \log \det \left(G_{0,\chi}^{-1}-\tilde{\Sigma}\right) - \frac{N}{2} \log \det \left(G_{0,\psi}^{-1}\right)
          \\& +
           \int_\mathcal{C}ds\,ds'\  \left[-\frac{N V^2}{2} \tilde{G} g F +\frac{N}{2} \left( \tilde{G} \tilde{\Sigma}-\frac{J^2\tilde{G}^4}{4}F \right) \right].
           \end{aligned}
           \end{equation}
where we've introduced the factor $F(s,s')\equiv f(s)f(s')$. We have $G_{0,\chi}(s,s')=\frac{1}{2}\text{sgn}(s-s')$ if both $s,s'$ lie on the same contour in fig. \ref{fig:fig2}(b) and otherwise zero. $G_{0,\psi}(x-x',s,s')=\left<T_\mathcal{C}\psi(x,s)\psi(x',s)\right>$ is the Green's function for the bath fermion $\psi$, without coupling to the SYK system. Since $\chi$ only couples to bath $\psi$ at $x=0$, only $g(s,s')\equiv G_{0,\psi}(0,s,s')\Lambda$ appears in the second line of \eqref{Seff}. For completeness, the explicit expression for $g(s,s')$ is given in Appendix \ref{A}.

We calculate the large $N$ leading order result of $S_{\chi_L,\chi_R}^{(2)}$ by doing saddle point approximations to the above path integral. The saddle point equations are
     \begin{align}\label{sd}
           G^{-1}&=G_{0,\chi}^{-1}-\Sigma,\ \ \ \ \  \Sigma=(J^2G^3+V^2g )F.
       \end{align}
This set of equations can be solved using iteration numerically. Generally, there could be several different solutions to the saddle point equation and the dominating solution in the large $N$ limit is determined by comparing the action. The on-shell action $I_{\mathcal{C}}$ can be written as
    \begin{equation}
    \begin{aligned}
      \frac{I_{\mathcal{C}}}{N}&= - \frac{1}{2} \log \det \left(G_{0,\chi}^{-1}-\Sigma\right) - \frac{1}{2} \log \det \left(G_{0,\psi}^{-1}\right) 
    +   \int_\mathcal{C}ds\,ds'\  \frac{3J^2G^4}{8} .
     \end{aligned}
    \end{equation}
Here the $\log\det$ term for $\psi$ does not depend on the saddle point solution $G_\chi$ and cancels with the normalization $Z(\beta)$.

\section{Numerical results} 
\label{sec:numerics}

\begin{comment}

First we need to explain how we parametrize the time in numerics. In fig. \ref{para}, the inner circle is the system ($\chi$), while the outer circle is the bath ($\psi$). Green and red colors represent forward and backward time evolution. To make the figure look clean, I used different shapes to denote how we should identify the points. Now, the way that we parametrize the time in numerics is to start from $0$ in the figure, and first parametrize the left circle and then parametrize the right circle, both clockwisely. We also show the free Green's function for $\chi$ under such parametrization of time in \ref{para} (b). 

\begin{figure}[h!]
  \center
  \includegraphics[width=1\columnwidth]{para.pdf}
  \caption{(a). How we parametrize the time in numerics. (b). The free Green's function for $\chi$ system.} \label{para}  
 \end{figure}

\end{comment}

We numerically solve the self-consisitent equation \eqref{sd} and the results of $S_{\chi_L,\chi_R}^{(2)}(t)$ for $\beta J=4$ and $\Lambda=5J$ are shown in Fig. \ref{num}. The result is qualitatively different for large and small coupling $V^2/J$:

\begin{figure}[h!]
      \center   
      \includegraphics[width=1\columnwidth]{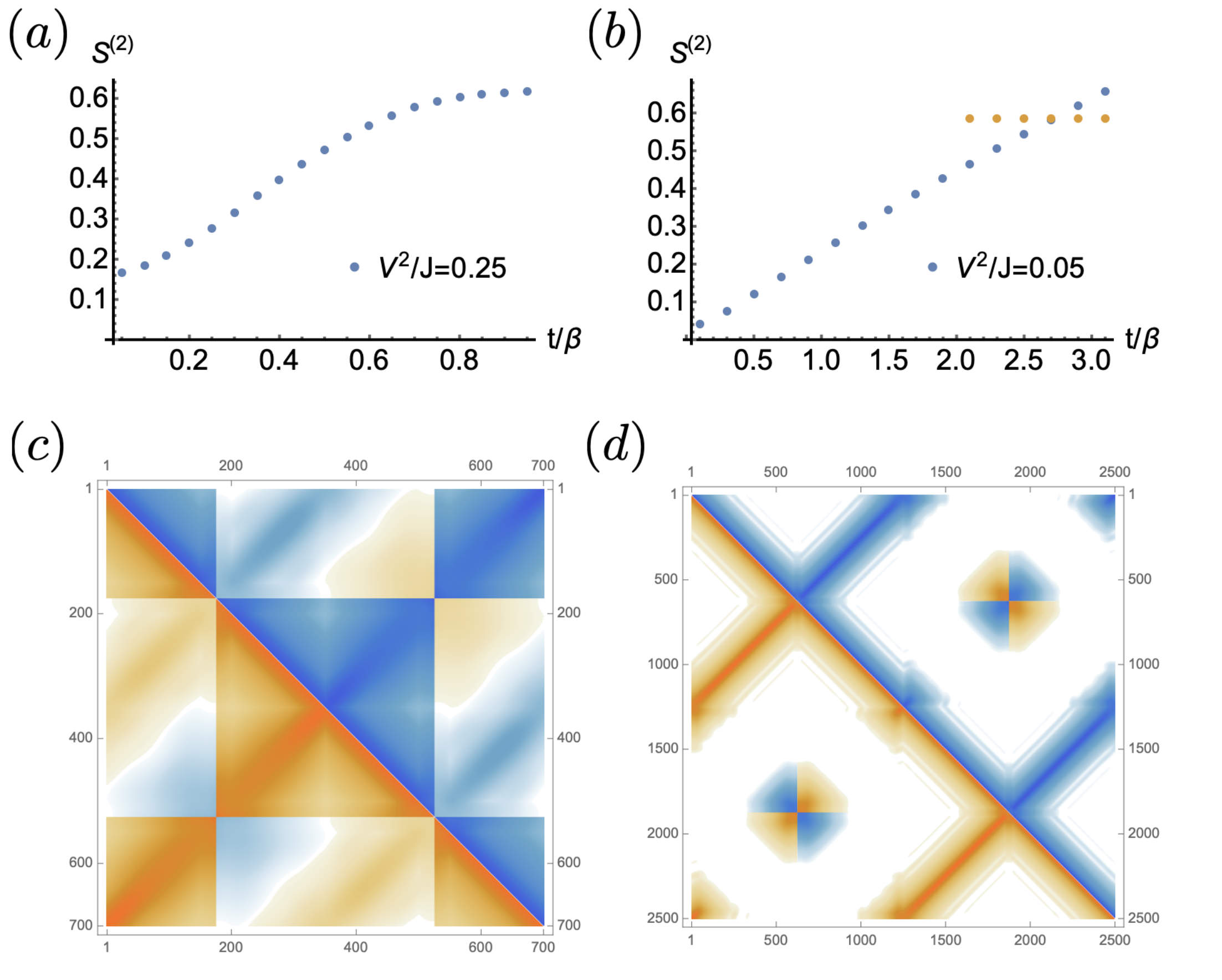}
        \caption{(a) Numerical result for $V^2/J=0.25$, $\beta J=4$ and $\Lambda=5J$. The entorpy is a smooth function of time. (b) Numerical result for $V^2/J=0.05$, $\beta J=4$ and $\Lambda=5J$. There is a first order transition of the entropy. (c) The real part of Green's function $G(s,s')$ corresponding to the short-time saddle in (b). Here we take $t/\beta=1.5$ as an example. Orange/Blue means positive/negative while their darkness indicates magnitude. The numbers on the axes correspond to the discretization of parameter $s$ in numerics. (d) The real part of Green's function $G(s,s')$ corresponding to the long-time saddle in (b). Here we take $t/\beta=6$ as an example. Here and in latter figures we have removed tiny matrix elements $|G(s,s')|<10^{-3}$ in the plot to make the plot clearer.} 
        \label{num}   
         \end{figure}

1. For the large coupling case $V^2/J=0.25$, as shown in Fig. \ref{num}(a), the entropy is a smooth function of time.

2. For small coupling $V^2/J=0.05$ case in (b), there is a first-order transition between two different saddle point solutions. The entropy initially grows almost linearly in time, and then switches to be almost time-independent, governed by a different saddle point. The two saddle points coexist for a finite time interval, which can be reached by choosing different initial conditions for the iteration. The transition time is the analog of Page time in evaporating black holes \cite{almheiri2019islands}.

We will study the solutions in more detail in latter sections. Here we just mention several properties of the solutions:

Firstly, the short-time solution for the small $V^2/J$ case is almost replica diagonal: the correlation between two SYK Majorana operators on different solid contours in Fig. \ref{fig:fig2}(b) is small. On the other hand, the long-time saddle is highly non-diagonal. The anti-diagonal peak for the long-time saddle is from the approximate cancellation between the forward and backward evolution for two halves of the $\chi$ contour interacting with the same $\psi$ system. There is a change of the paring between the forward and the backward evolution, which has been found in \cite{Maldacena:2018lmt} for a coupled SYK system. Note that the forward evolution and the backward evolution on the same solid contour of Fig. \ref{fig:fig2}(b) could not cancel exactly since they interact with different $\psi$ systems.

Secondly, for large $V^2/J$, the off-diagonal terms are of the same order as the diagonal terms. The short-time solution described above is smoothly connected to the long-time solution. This is similar to the equilibrium problem of coupled SYK model \cite{Maldacena:2018lmt}, where for a small coupling there is a first order transition, while for a large coupling the free energy is smooth.

Thirdly, the entropy for the long-time solution is almost constant and close to $2S^{(2)}_{th}$, where $S^{(2)}_{th}$ is the second R\'{e}nyi entropy between $\chi$ and $\psi$ in a thermal density matrix with the Hamiltonian in (\ref{model1H}). 
        
\section{The short-time solution}\label{analyticshort}

In this subsection, we discuss the analytic calculation of the R\'{e}nyi entropy. Without turning on interaction between the system and the bath in the real-time evolution, the entropy $S^{(2)}$ is time-independent. As a result, focusing on the time dependence of $S^{(2)}$, we perform a perturbative calculation in $V^2$ starting from the $V=0$ replica diagonal solution for short-time. The calculation is similar to \cite{gu2017spread}, where one first calculates the action in Euclidean time and then continue the result to Lorentzian time. 

\begin{figure}[t]
  \center
  \includegraphics[width=0.75\columnwidth]{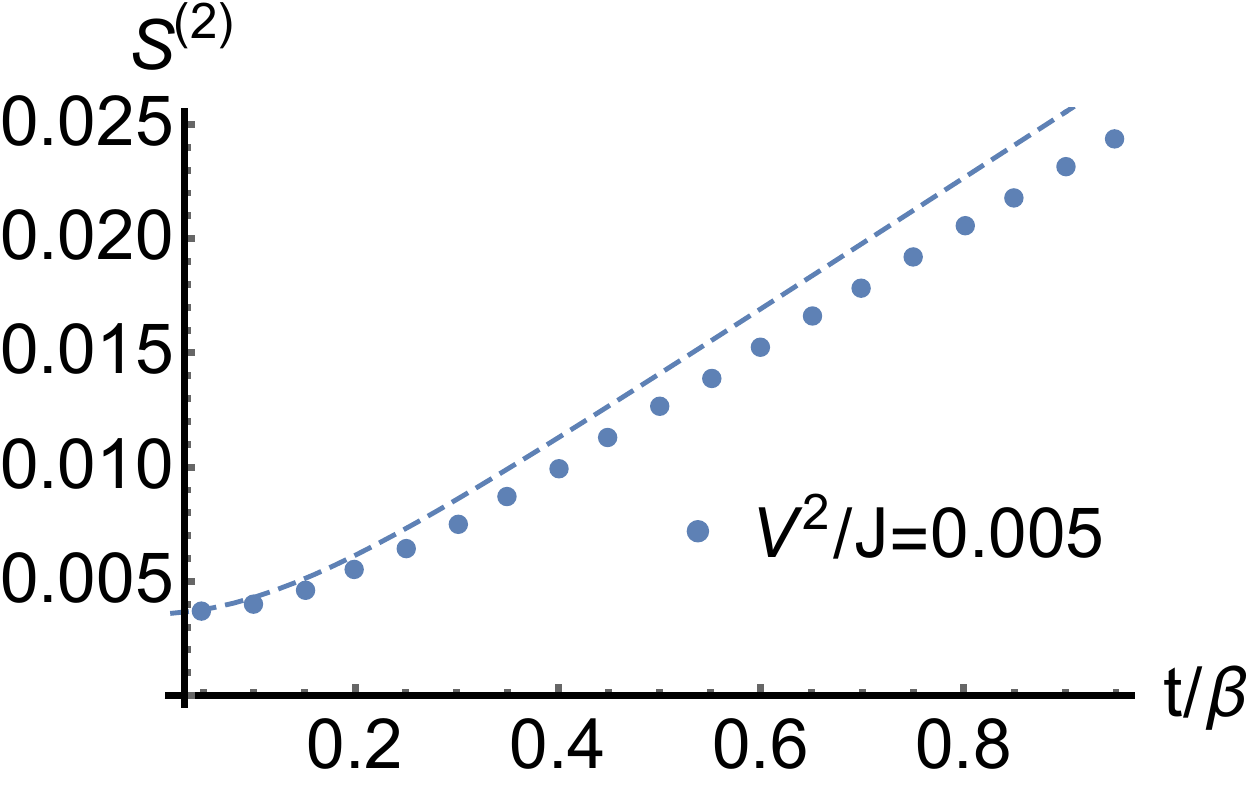}
  \caption{Comparison of the analytic formula \eqref{resshort} with numerics for $V^2/J = 0.005$, $\beta J = 4$ and $\Lambda =5J$.}
  \label{pert}  
 \end{figure}

The conformal limit solution for the $\chi$ system is
        \begin{equation}
            G(\tau_1,\tau_2) = G_c(\tau_1-\tau_2)= \frac{1}{(4\pi J^2 )^{1/4}} \left( \frac{\pi}{\beta \sin \frac{\pi (\tau_1 - \tau_2)}{\beta} } \right)^{\frac{1}{2}}\label{Gchi}
        \end{equation}
Similarly, at the low-temperature limit $\Lambda \beta \rightarrow \infty$, we expect $G_{0,\psi}(0,\tau_1,\tau_2)$ to be dominated by low energy modes. We can use the conformal two-point function with scaling dimension $\Delta=1/2$ for the bath:
 \begin{equation}
    g(\tau_1,\tau_2)=\frac{1}{\pi}\left(\frac{\pi}{\beta \sin \frac{\pi  (\tau_1-\tau_2) }{\beta }}\right).\label{Gpsi}
 \end{equation}
To calculate the change of entropy, we rewrite the $n$-replica action as
 \begin{equation}\label{effa}
        \begin{aligned}
           \frac{I^{(n)}}{n}&= - \frac{N}{2} \log \det \left(\partial_\tau-\Sigma\right) - \frac{N}{2} \log \det \left(G_{0,\psi}^{-1}\right)
          \\& +
           \int_0^\beta d\tau_1\int_0^\beta d\tau_2\  \left[-\frac{N V^2}{2} G g +\frac{N}{2} \left( G \Sigma-\frac{J^2G^4}{4} \right) \right]\\&+\int_\epsilon^{\tau_0-\epsilon}d\tau_1\int_{\tau_0+\epsilon}^{\beta-\epsilon}d\tau_2\ N V^2 G g. \\
         & = \frac{1}{n} S_0 + \frac{1}{n} \Delta S,
        \end{aligned}
        \end{equation}
        \begin{equation}
            \frac{(n-1)S^{(n)}_{\chi_L, \chi_R}}{N}=\frac{I^{(n)} + n \log Z}{N} =  \frac{\Delta S}{N} =  n V^2 \int_\epsilon^{\tau_0-\epsilon}d\tau_1\int_{\tau_0+\epsilon}^{\beta-\epsilon}d\tau_2\ G(\tau_1,\tau_2) g(\tau_1,\tau_2), \label{action short}
        \end{equation}
 where $\epsilon$ is a small UV regulator. We take $\tau_0=\frac{\beta}{2}-2it$ in the end. The details are provided in Appendix \ref{B}. For time $t\gg\beta$, the entropy shows a linear growing behavior:
\begin{align}\label{entropy linear}
\frac{S^{(2)}_{\chi_L, \chi_R} (t)}{N}=\frac{8 V^2 \Gamma \left(\frac{3}{4}\right)^2 }{\pi ^{5/4}}\sqrt{\frac{\beta }{J}}\frac{t}{\beta} + \textrm{const}.
\end{align}
Here in the slope of the linear growth, there is an additional factor $\sqrt{\beta}$ compared to the results in \cite{gu2017spread,penington2019replica}. This is due to the fact that the coupling $V$ is relevant with scaling dimension $1/4$, whose effect becomes larger when we lower the temperature. In Fig. \ref{pert}, we compare the analytic formula \eqref{resshort} with the numerics for $V^2/J = 0.005$, $\beta J = 4$ and $\Lambda =5J$. There we have chosen the constant piece in (\ref{entropy linear}) to match the $t=0$ numerical result. 

The linear entropy growth in \eqref{entropy linear} can not persist to time much long than $t\sim \frac{\sqrt{\beta J}}{V^2} $, since the entropy cannot exceed the thermal entropy at the same energy density. % where the second Renyi mutual information between $\chi_L$ and $\chi_R$ would become negative. 
At this time scale, a first order transition occurs for small $V^2/J$. Note that here the Page time is finite at $N\rightarrow \infty$, because the central charge of the bath is proportional to $N$ and there is an extensive entropy flow between the bath and the system. This is also consistent with the gravity analysis in \cite{almheiri2019islands} with $c \propto N$.

\begin{figure}[t]
  \center
  
\subfigure[]  
{  
\begin{tikzpicture}[thick,scale = 0.75]
 \draw (1,0.2) arc(0:180:1 and 1);
 \draw (1,-0.2) arc(0:-180:1 and 1);
  \draw (-1,-0.2) -- (-2.5,-0.2);
   \draw (-1,0.2) -- (-2.5,0.2);

   \draw (1.2,0.2) arc(0:170:1.2 and 1.2);
   \draw (1.2,-0.2) arc(0:-170:1.2 and 1.2);
   \draw[dashed] (1.2,0.2) --(1.2,-0.2) ;
    \draw (-1.18,-0.4) -- (-2.5,-0.4);
     \draw (-1.18,0.4) -- (-2.5,0.4);
     \draw[dashed] (-2.5,0.4) arc (90:270:0.4 and 0.4);
     \draw[dotted] (0.866,0.7) -- (1.039,0.8);
       \draw[dotted] (0.866,-0.7) -- (1.039,-0.8);
      \draw[dotted] (0.5,1.066) -- (0.6,1.239);
        \draw[dotted] (0.5,-1.066) -- (0.6,-1.239);
           \draw[dotted] (0,1.2) -- (0,1.4);
              \draw[dotted] (0,-1.2) -- (0,-1.4);
          \draw[dotted] (-0.5,1.066) -- (-0.6,1.239);
        \draw[dotted] (-0.5,-1.066) -- (-0.6,-1.239);
         \draw[dotted] (-0.866,0.7) -- (-1.039,0.8);
           \draw[dotted] (-0.866,-0.7) -- (-1.039,-0.8);
           \draw[dotted] (-1.4,0.2) -- (-1.4,0.4);
             \draw[dotted] (-1.4,-0.2) -- (-1.4,-0.4);
                     \draw[dotted] (-1.8,0.2) -- (-1.8,0.4);
             \draw[dotted] (-1.8,-0.2) -- (-1.8,-0.4);
              \draw[dotted] (-2.2,0.2) -- (-2.2,0.4);
             \draw[dotted] (-2.2,-0.2) -- (-2.2,-0.4);
      \draw (0.5,-0.8) node{$\chi$};
       \draw (0.7,-1.4) node{$\psi$};
 
 \begin{scope}[shift={(0,3)}]
 \draw (1,0.2) arc(0:180:1 and 1);
 \draw (1,-0.2) arc(0:-180:1 and 1);
  \draw (-1,-0.2) -- (-2.5,-0.2);
   \draw (-1,0.2) -- (-2.5,0.2);

   \draw (1.2,0.2) arc(0:170:1.2 and 1.2);
   \draw (1.2,-0.2) arc(0:-170:1.2 and 1.2);
   \draw[dashed] (1.2,0.2) --(1.2,-0.2) ;
    \draw (-1.18,-0.4) -- (-2.5,-0.4);
     \draw (-1.18,0.4) -- (-2.5,0.4);
     \draw[dashed] (-2.5,0.4) arc (90:270:0.4 and 0.4);
     \draw[dotted] (0.866,0.7) -- (1.039,0.8);
       \draw[dotted] (0.866,-0.7) -- (1.039,-0.8);
      \draw[dotted] (0.5,1.066) -- (0.6,1.239);
        \draw[dotted] (0.5,-1.066) -- (0.6,-1.239);
           \draw[dotted] (0,1.2) -- (0,1.4);
              \draw[dotted] (0,-1.2) -- (0,-1.4);
          \draw[dotted] (-0.5,1.066) -- (-0.6,1.239);
        \draw[dotted] (-0.5,-1.066) -- (-0.6,-1.239);
         \draw[dotted] (-0.866,0.7) -- (-1.039,0.8);
           \draw[dotted] (-0.866,-0.7) -- (-1.039,-0.8);
           \draw[dotted] (-1.4,0.2) -- (-1.4,0.4);
             \draw[dotted] (-1.4,-0.2) -- (-1.4,-0.4);
                     \draw[dotted] (-1.8,0.2) -- (-1.8,0.4);
             \draw[dotted] (-1.8,-0.2) -- (-1.8,-0.4);
              \draw[dotted] (-2.2,0.2) -- (-2.2,0.4);
             \draw[dotted] (-2.2,-0.2) -- (-2.2,-0.4);
      \draw (0.5,-0.8) node{$\chi$};
       \draw (0.7,-1.4) node{$\psi$};
       
\end{scope}
\draw[dashed] (-2.5,3+0.2) arc (90:270:1 and 1.7);
\draw[dashed] (-2.5,3-0.2) arc (90:270:0.5 and 1.3);
\draw[dashed] (1,3+0.2) arc (90:-90:1 and 1.7);
\draw[dashed] (1,3-0.2) arc (90:-90:0.5 and 1.3);

 \draw (-2.5,3+0.6-0.03) node[circle,fill,scale = 0.1,above]{A};
 \draw[->] (-2.5, 3+0.6) -- (-1.5, 3+0.6) ;
 \draw (-2.3,3+0.9) node{$s=0$} ;
 
 \draw (-2.5,0.6-0.03) node[circle,fill,scale = 0.1,above]{A};
 \draw[->] (-2.5, 0.6) -- (-1.5, 0.6) ;
 \draw (-2.1,0.9) node{$s=\beta + 4t$} ;
 
\end{tikzpicture}

  \includegraphics[width=0.7\columnwidth]{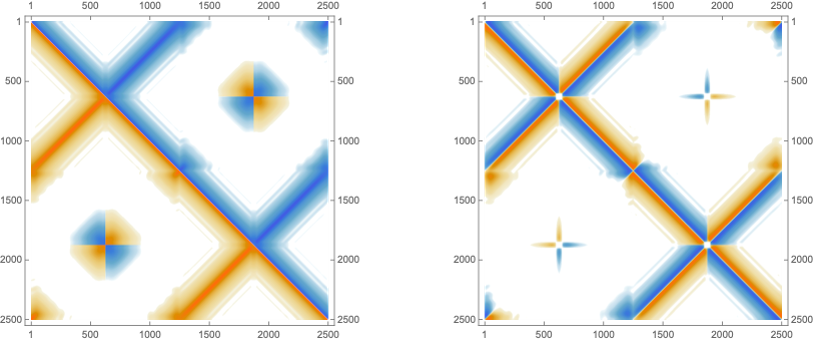}

}  

\subfigure[]  
{  
\begin{tikzpicture}[thick,scale = 0.75]
 \draw (1,0.2) arc(0:180:1 and 1);
 \draw (1,-0.2) arc(0:-180:1 and 1);
  \draw (-1,-0.2) -- (-2.5,-0.2);
   \draw (-1,0.2) -- (-2.5,0.2);

   \draw (1.2,0.2) arc(0:170:1.2 and 1.2);
   \draw (1.2,-0.2) arc(0:-170:1.2 and 1.2);
   \draw (1.2,0.2) --(1.2,-0.2) ;
    \draw (-1.18,-0.4) -- (-2.5,-0.4);
     \draw (-1.18,0.4) -- (-2.5,0.4);
     \draw[dashed] (-2.5,0.4) arc (90:270:0.4 and 0.4);
     \draw[dotted] (0.866,0.7) -- (1.039,0.8);
       \draw[dotted] (0.866,-0.7) -- (1.039,-0.8);
      \draw[dotted] (0.5,1.066) -- (0.6,1.239);
        \draw[dotted] (0.5,-1.066) -- (0.6,-1.239);
           \draw[dotted] (0,1.2) -- (0,1.4);
              \draw[dotted] (0,-1.2) -- (0,-1.4);
          \draw[dotted] (-0.5,1.066) -- (-0.6,1.239);
        \draw[dotted] (-0.5,-1.066) -- (-0.6,-1.239);
         \draw[dotted] (-0.866,0.7) -- (-1.039,0.8);
           \draw[dotted] (-0.866,-0.7) -- (-1.039,-0.8);
           \draw[dotted] (-1.4,0.2) -- (-1.4,0.4);
             \draw[dotted] (-1.4,-0.2) -- (-1.4,-0.4);
                     \draw[dotted] (-1.8,0.2) -- (-1.8,0.4);
             \draw[dotted] (-1.8,-0.2) -- (-1.8,-0.4);
              \draw[dotted] (-2.2,0.2) -- (-2.2,0.4);
             \draw[dotted] (-2.2,-0.2) -- (-2.2,-0.4);
      \draw (0.5,-0.8) node{$\chi$};
       \draw (0.7,-1.4) node{$\psi$};
 
 \begin{scope}[shift={(0,3)}]
 \draw (1,0.2) arc(0:180:1 and 1);
 \draw (1,-0.2) arc(0:-180:1 and 1);
  \draw (-1,-0.2) -- (-2.5,-0.2);
   \draw (-1,0.2) -- (-2.5,0.2);

   \draw (1.2,0.2) arc(0:170:1.2 and 1.2);
   \draw (1.2,-0.2) arc(0:-170:1.2 and 1.2);
   \draw (1.2,0.2) --(1.2,-0.2) ;
    \draw (-1.18,-0.4) -- (-2.5,-0.4);
     \draw (-1.18,0.4) -- (-2.5,0.4);
     \draw[dashed] (-2.5,0.4) arc (90:270:0.4 and 0.4);
     \draw[dotted] (0.866,0.7) -- (1.039,0.8);
       \draw[dotted] (0.866,-0.7) -- (1.039,-0.8);
      \draw[dotted] (0.5,1.066) -- (0.6,1.239);
        \draw[dotted] (0.5,-1.066) -- (0.6,-1.239);
           \draw[dotted] (0,1.2) -- (0,1.4);
              \draw[dotted] (0,-1.2) -- (0,-1.4);
          \draw[dotted] (-0.5,1.066) -- (-0.6,1.239);
        \draw[dotted] (-0.5,-1.066) -- (-0.6,-1.239);
         \draw[dotted] (-0.866,0.7) -- (-1.039,0.8);
           \draw[dotted] (-0.866,-0.7) -- (-1.039,-0.8);
           \draw[dotted] (-1.4,0.2) -- (-1.4,0.4);
             \draw[dotted] (-1.4,-0.2) -- (-1.4,-0.4);
                     \draw[dotted] (-1.8,0.2) -- (-1.8,0.4);
             \draw[dotted] (-1.8,-0.2) -- (-1.8,-0.4);
              \draw[dotted] (-2.2,0.2) -- (-2.2,0.4);
             \draw[dotted] (-2.2,-0.2) -- (-2.2,-0.4);
      \draw (0.5,-0.8) node{$\chi$};
       \draw (0.7,-1.4) node{$\psi$};
       
\end{scope}
\draw[dashed] (-2.5,3+0.2) arc (90:270:0.2 and 0.2);
\draw[dashed] (-2.5,0.2) arc (90:270:0.2 and 0.2);
\draw[solid] (1,3+0.2) arc (90:-90:0.01 and 0.2);
\draw[solid] (1,0.2) arc (90:-90:0.01 and 0.2);

 \draw (-2.5,3+0.6-0.03) node[circle,fill,scale = 0.1,above]{A};
 \draw[->] (-2.5, 3+0.6) -- (-1.5, 3+0.6) ;
 \draw (-2.3,3+0.9) node{$s=0$} ;
 
 \draw (-2.5,0.6-0.03) node[circle,fill,scale = 0.1,above]{A};
 \draw[->] (-2.5, 0.6) -- (-1.5, 0.6) ;
 \draw (-2.1,0.9) node{$s=\beta + 4t$} ;
 
\end{tikzpicture}

 ~ ~ \includegraphics[width=0.7\columnwidth]{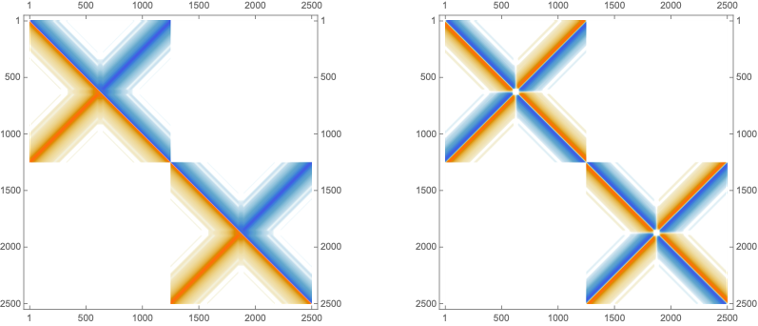}

}

  \caption{(a) The long-time solution with $t=6\beta, V^2/J=0.05$. The two matrix plots are the real part and the imaginary part respectively (same in (b)). (b) The trivial solution that we do not insert any twist operators. } 
  \label{long}  
 \end{figure}
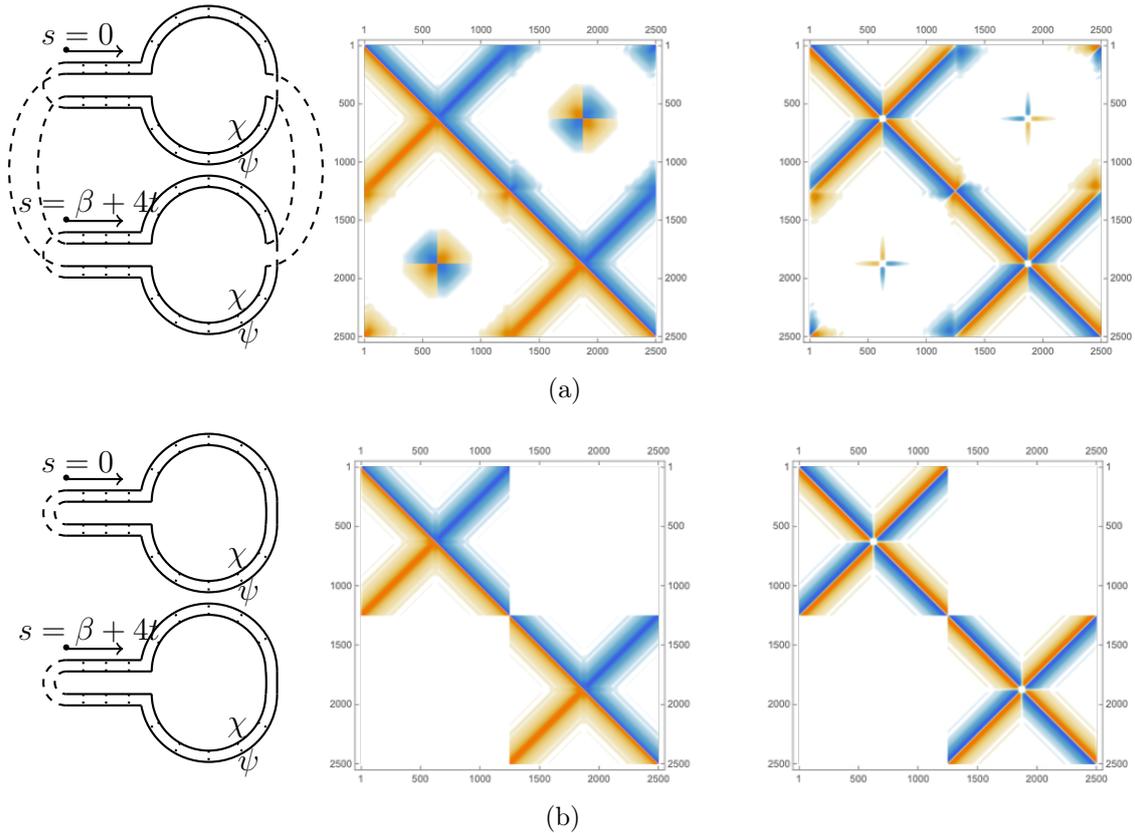

\section{The long-time solution}\label{sec:analyticlong}

In this section, we discuss the structure of the long-time solution and give arguments for the long-time saturation value.
        
To gain some insight of the problem, in fig. \ref{long}(a), we show how the long-time solution of $G$ looks like (it is similar to the one shown in fig. \ref{num}(d), but the time is longer, and we set the small elements to zero). In fig. \ref{long}(b), we show the solution with the same parameters, but on the contour without inserting any twist operators and thus being two disconnected circles. The important observation is that the solution of (a) is very similar to (b), with the only significant differences locating at the places near the twist operators. 

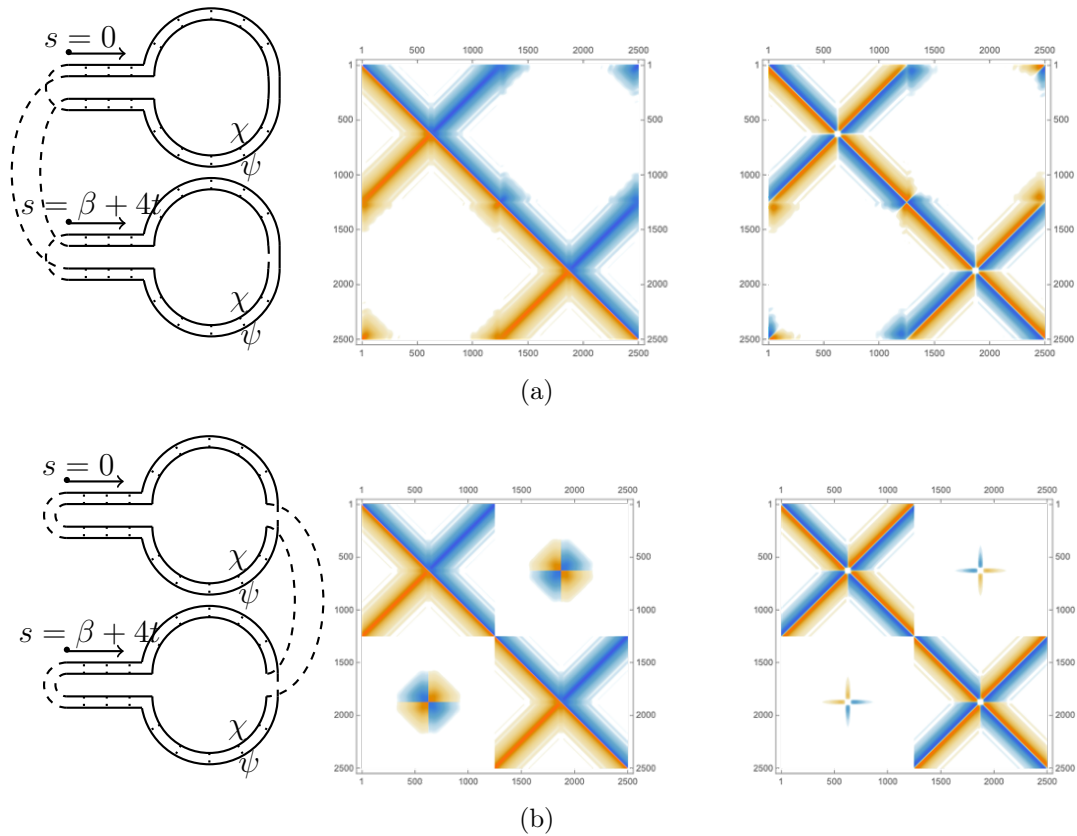
\begin{figure}[t]
  \center
  
\subfigure[]  
{  
\begin{tikzpicture}[thick,scale = 0.75]
 \draw (1,0.2) arc(0:180:1 and 1);
 \draw (1,-0.2) arc(0:-180:1 and 1);
  \draw (-1,-0.2) -- (-2.5,-0.2);
   \draw (-1,0.2) -- (-2.5,0.2);

   \draw (1.2,0.2) arc(0:170:1.2 and 1.2);
   \draw (1.2,-0.2) arc(0:-170:1.2 and 1.2);
   \draw (1.2,0.2) --(1.2,-0.2) ;
    \draw (-1.18,-0.4) -- (-2.5,-0.4);
     \draw (-1.18,0.4) -- (-2.5,0.4);
     \draw[dashed] (-2.5,0.4) arc (90:270:0.4 and 0.4);
     \draw[dotted] (0.866,0.7) -- (1.039,0.8);
       \draw[dotted] (0.866,-0.7) -- (1.039,-0.8);
      \draw[dotted] (0.5,1.066) -- (0.6,1.239);
        \draw[dotted] (0.5,-1.066) -- (0.6,-1.239);
           \draw[dotted] (0,1.2) -- (0,1.4);
              \draw[dotted] (0,-1.2) -- (0,-1.4);
          \draw[dotted] (-0.5,1.066) -- (-0.6,1.239);
        \draw[dotted] (-0.5,-1.066) -- (-0.6,-1.239);
         \draw[dotted] (-0.866,0.7) -- (-1.039,0.8);
           \draw[dotted] (-0.866,-0.7) -- (-1.039,-0.8);
           \draw[dotted] (-1.4,0.2) -- (-1.4,0.4);
             \draw[dotted] (-1.4,-0.2) -- (-1.4,-0.4);
                     \draw[dotted] (-1.8,0.2) -- (-1.8,0.4);
             \draw[dotted] (-1.8,-0.2) -- (-1.8,-0.4);
              \draw[dotted] (-2.2,0.2) -- (-2.2,0.4);
             \draw[dotted] (-2.2,-0.2) -- (-2.2,-0.4);
      \draw (0.5,-0.8) node{$\chi$};
       \draw (0.7,-1.4) node{$\psi$};
 
 \begin{scope}[shift={(0,3)}]
 \draw (1,0.2) arc(0:180:1 and 1);
 \draw (1,-0.2) arc(0:-180:1 and 1);
  \draw (-1,-0.2) -- (-2.5,-0.2);
   \draw (-1,0.2) -- (-2.5,0.2);

   \draw (1.2,0.2) arc(0:170:1.2 and 1.2);
   \draw (1.2,-0.2) arc(0:-170:1.2 and 1.2);
   \draw (1.2,0.2) --(1.2,-0.2) ;
    \draw (-1.18,-0.4) -- (-2.5,-0.4);
     \draw (-1.18,0.4) -- (-2.5,0.4);
     \draw[dashed] (-2.5,0.4) arc (90:270:0.4 and 0.4);
     \draw[dotted] (0.866,0.7) -- (1.039,0.8);
       \draw[dotted] (0.866,-0.7) -- (1.039,-0.8);
      \draw[dotted] (0.5,1.066) -- (0.6,1.239);
        \draw[dotted] (0.5,-1.066) -- (0.6,-1.239);
           \draw[dotted] (0,1.2) -- (0,1.4);
              \draw[dotted] (0,-1.2) -- (0,-1.4);
          \draw[dotted] (-0.5,1.066) -- (-0.6,1.239);
        \draw[dotted] (-0.5,-1.066) -- (-0.6,-1.239);
         \draw[dotted] (-0.866,0.7) -- (-1.039,0.8);
           \draw[dotted] (-0.866,-0.7) -- (-1.039,-0.8);
           \draw[dotted] (-1.4,0.2) -- (-1.4,0.4);
             \draw[dotted] (-1.4,-0.2) -- (-1.4,-0.4);
                     \draw[dotted] (-1.8,0.2) -- (-1.8,0.4);
             \draw[dotted] (-1.8,-0.2) -- (-1.8,-0.4);
              \draw[dotted] (-2.2,0.2) -- (-2.2,0.4);
             \draw[dotted] (-2.2,-0.2) -- (-2.2,-0.4);
      \draw (0.5,-0.8) node{$\chi$};
       \draw (0.7,-1.4) node{$\psi$};
       
\end{scope}
\draw[dashed] (-2.5,3+0.2) arc (90:270:1 and 1.7);
\draw[dashed] (-2.5,3-0.2) arc (90:270:0.5 and 1.3);
\draw[solid] (1,3+0.2) arc (90:-90:0.01 and 0.2);
\draw[dashed] (1,0.2) arc (90:-90:0.01 and 0.2);

 \draw (-2.5,3+0.6-0.03) node[circle,fill,scale = 0.1,above]{A};
 \draw[->] (-2.5, 3+0.6) -- (-1.5, 3+0.6) ;
 \draw (-2.3,3+0.9) node{$s=0$} ;
 
 \draw (-2.5,0.6-0.03) node[circle,fill,scale = 0.1,above]{A};
 \draw[->] (-2.5, 0.6) -- (-1.5, 0.6) ;
 \draw (-2.1,0.9) node{$s=\beta + 4t$} ;
 
\end{tikzpicture}

  ~ ~ \includegraphics[width=0.65\columnwidth]{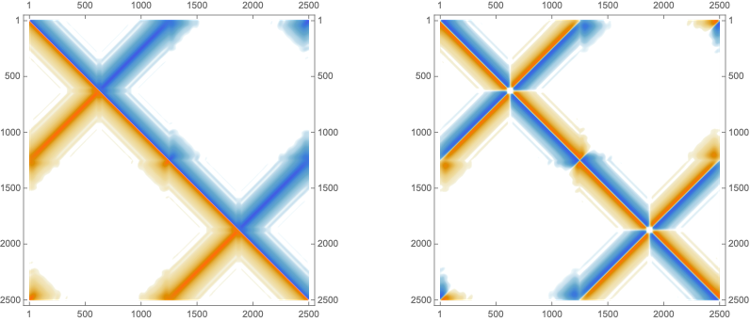}

}  

\subfigure[]  
{  
\begin{tikzpicture}[thick,scale = 0.75]
 \draw (1,0.2) arc(0:180:1 and 1);
 \draw (1,-0.2) arc(0:-180:1 and 1);
  \draw (-1,-0.2) -- (-2.5,-0.2);
   \draw (-1,0.2) -- (-2.5,0.2);

   \draw (1.2,0.2) arc(0:170:1.2 and 1.2);
   \draw (1.2,-0.2) arc(0:-170:1.2 and 1.2);
   \draw[dashed] (1.2,0.2) --(1.2,-0.2) ;
    \draw (-1.18,-0.4) -- (-2.5,-0.4);
     \draw (-1.18,0.4) -- (-2.5,0.4);
     \draw[dashed] (-2.5,0.4) arc (90:270:0.4 and 0.4);
     \draw[dotted] (0.866,0.7) -- (1.039,0.8);
       \draw[dotted] (0.866,-0.7) -- (1.039,-0.8);
      \draw[dotted] (0.5,1.066) -- (0.6,1.239);
        \draw[dotted] (0.5,-1.066) -- (0.6,-1.239);
           \draw[dotted] (0,1.2) -- (0,1.4);
              \draw[dotted] (0,-1.2) -- (0,-1.4);
          \draw[dotted] (-0.5,1.066) -- (-0.6,1.239);
        \draw[dotted] (-0.5,-1.066) -- (-0.6,-1.239);
         \draw[dotted] (-0.866,0.7) -- (-1.039,0.8);
           \draw[dotted] (-0.866,-0.7) -- (-1.039,-0.8);
           \draw[dotted] (-1.4,0.2) -- (-1.4,0.4);
             \draw[dotted] (-1.4,-0.2) -- (-1.4,-0.4);
                     \draw[dotted] (-1.8,0.2) -- (-1.8,0.4);
             \draw[dotted] (-1.8,-0.2) -- (-1.8,-0.4);
              \draw[dotted] (-2.2,0.2) -- (-2.2,0.4);
             \draw[dotted] (-2.2,-0.2) -- (-2.2,-0.4);
      \draw (0.5,-0.8) node{$\chi$};
       \draw (0.7,-1.4) node{$\psi$};
 
 \begin{scope}[shift={(0,3)}]
 \draw (1,0.2) arc(0:180:1 and 1);
 \draw (1,-0.2) arc(0:-180:1 and 1);
  \draw (-1,-0.2) -- (-2.5,-0.2);
   \draw (-1,0.2) -- (-2.5,0.2);

   \draw (1.2,0.2) arc(0:170:1.2 and 1.2);
   \draw (1.2,-0.2) arc(0:-170:1.2 and 1.2);
   \draw[dashed] (1.2,0.2) --(1.2,-0.2) ;
    \draw (-1.18,-0.4) -- (-2.5,-0.4);
     \draw (-1.18,0.4) -- (-2.5,0.4);
     \draw[dashed] (-2.5,0.4) arc (90:270:0.4 and 0.4);
     \draw[dotted] (0.866,0.7) -- (1.039,0.8);
       \draw[dotted] (0.866,-0.7) -- (1.039,-0.8);
      \draw[dotted] (0.5,1.066) -- (0.6,1.239);
        \draw[dotted] (0.5,-1.066) -- (0.6,-1.239);
           \draw[dotted] (0,1.2) -- (0,1.4);
              \draw[dotted] (0,-1.2) -- (0,-1.4);
          \draw[dotted] (-0.5,1.066) -- (-0.6,1.239);
        \draw[dotted] (-0.5,-1.066) -- (-0.6,-1.239);
         \draw[dotted] (-0.866,0.7) -- (-1.039,0.8);
           \draw[dotted] (-0.866,-0.7) -- (-1.039,-0.8);
           \draw[dotted] (-1.4,0.2) -- (-1.4,0.4);
             \draw[dotted] (-1.4,-0.2) -- (-1.4,-0.4);
                     \draw[dotted] (-1.8,0.2) -- (-1.8,0.4);
             \draw[dotted] (-1.8,-0.2) -- (-1.8,-0.4);
              \draw[dotted] (-2.2,0.2) -- (-2.2,0.4);
             \draw[dotted] (-2.2,-0.2) -- (-2.2,-0.4);
      \draw (0.5,-0.8) node{$\chi$};
       \draw (0.7,-1.4) node{$\psi$};
       
\end{scope}
\draw[dashed] (-2.5,3+0.2) arc (90:270:0.2 and 0.2);
\draw[dashed] (-2.5,0.2) arc (90:270:0.2 and 0.2);
\draw[dashed] (1,3+0.2) arc (90:-90:1 and 1.7);
\draw[dashed] (1,3-0.2) arc (90:-90:0.5 and 1.3);

 \draw (-2.5,3+0.6-0.03) node[circle,fill,scale = 0.1,above]{A};
 \draw[->] (-2.5, 3+0.6) -- (-1.5, 3+0.6) ;
 \draw (-2.3,3+0.9) node{$s=0$} ;
 
 \draw (-2.5,0.6-0.03) node[circle,fill,scale = 0.1,above]{A};
 \draw[->] (-2.5, 0.6) -- (-1.5, 0.6) ;
 \draw (-2.1,0.9) node{$s=\beta + 4t$} ;
 
\end{tikzpicture}

  \includegraphics[width=0.65\columnwidth]{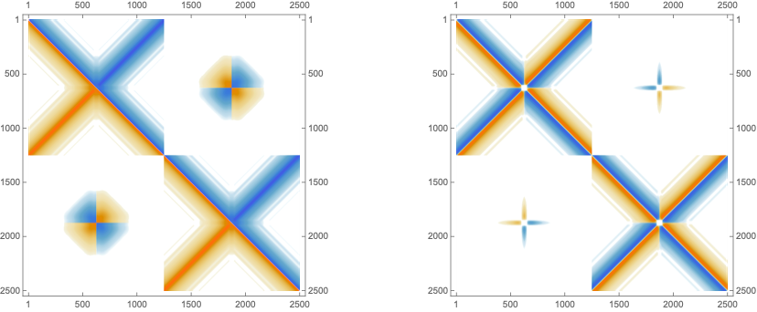}

}

  \caption{The solutions with only one insertion of the twist operator.} 
  \label{onetwist}  
 \end{figure}

What this tells us is that for the long-time solution, the backreaction of the twist operators is only local. Thus we expect the correlation function of the twist operators to be approximately factorized, i.e. 
\begin{equation}\label{factorizationtwist}
\langle T_L T_R \rangle \approx \langle T_L \rangle\langle T_R \rangle    .
\end{equation}
The solutions with only one insertion of twist operator are shown in fig. \ref{onetwist} (a)(b). A more careful argument for the relation between local back-reaction of twist operators on the on-shell solution and the factorization in (\ref{factorizationtwist}) is given in Appendix \ref{C}.

\begin{figure}[ht]
  \center
  \includegraphics[width=0.9\columnwidth]{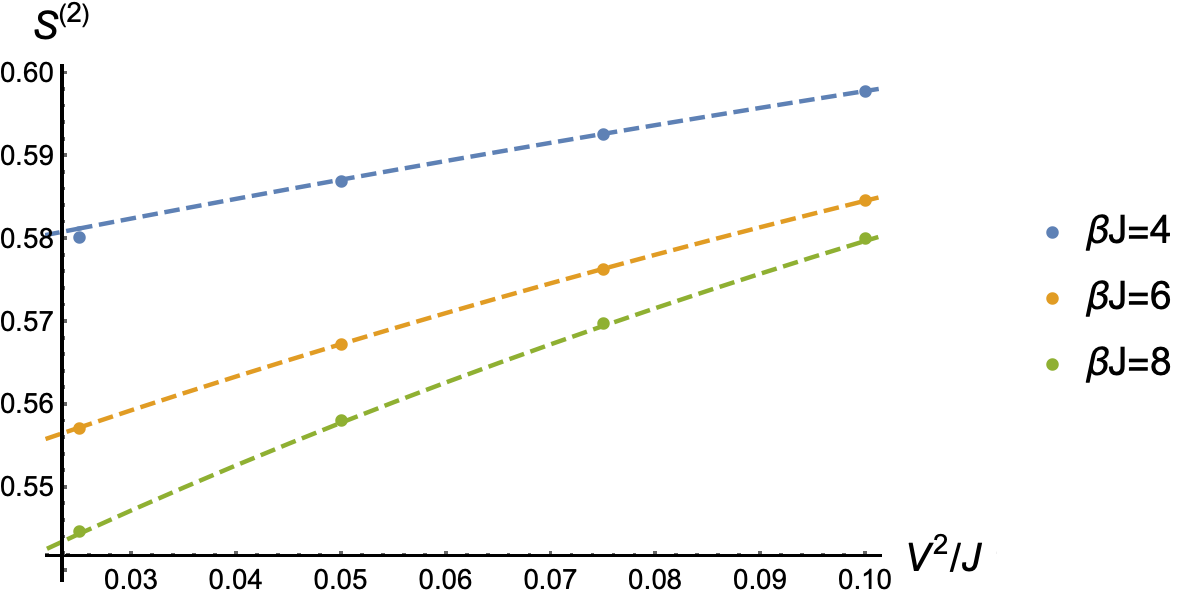}
  \caption{We compare the value of $S^{(2)}_{\chi_L,\chi_R}$ after the Page time (the dots) and twice the second R\'{e}nyi entropy $S_{th}^{(2)}$ between $\chi$ and $\psi$ in the thermal density matrix of a single system (the dashed lines).}
  \label{fig:saturation}  
 \end{figure}

For the computation of each one-point function $\langle T_{L/R}\rangle$, the other side of the system ($R/L$) is traced out within replicas. As a result, the one-point function of the twist operator computes the second R\'{e}nyi entropy between one copy of the system and the bath in a thermal state: one prepares the system $\chi$ and bath $\psi$ into a thermal ensemble with density matrix $\rho_{\text{th}}=\exp(-\beta H)/Z$, and then compute the second R\'{e}nyi entropy of the system $\chi$, denoted by $S_{th}^{(2)}$. Note that for the one-point function in a thermal ensemble, the time evolution is trivial since the density matrix commutes with the unitary evolution. Consequently, by the above argument, the saturation value of the second R\'{e}nyi entropy should be $S^{(2)}_{\chi_L,\chi_R}\simeq 2S^{(2)}_{th}$. This is also consistent with the gravity calculation in \cite{almheiri2019islands}. In fig. \ref{fig:saturation}, we see that this estimation agrees quite well with the numerical results.

\section{Information retrieval from the bath}\label{sec:retrieval}

By the gravity picture in \cite{penington2019entanglement,almheiri2019entropy,almheiri2019page,almheiri2019islands,almheiri2019replica,penington2019replica}, after the Page time, there is an island in the bulk which belongs to the entanglement wedge of the bath. This means that if one throws in a particle after the Page time and wait for a scrambling time, one should be able to tell the difference between the state with and without the particle by only accessing the state of the bath. In other words, the information carried by an infalling particle can be retrieved from the bath after a scrambling time. In comparison, a particle thrown in before Page time can only be retrieved after the Page time. From the quantum mechanical point of view, this suggests that an information initially scrambled in the SYK system should emerge in the bath after the Page time. In this section, we will study this phenomena in our model.

\begin{figure}[t]
  \center

\subfigure[]  
{    
  \begin{tikzpicture}[thick,scale = 1]

 \draw (1,0.2) arc(0:180:1 and 1);
 \draw (1,-0.2) arc(0:-180:1 and 1);
  \draw (-1,-0.2) -- (-2.5,-0.2);
   \draw (-1,0.2) -- (-2.5,0.2);

   \draw (1.2,0.2) arc(0:170:1.2 and 1.2);
   \draw (1.2,-0.2) arc(0:-170:1.2 and 1.2);
   
    \draw (-1.18,-0.4) -- (-2.5,-0.4);
     \draw (-1.18,0.4) -- (-2.5,0.4);
     
     \draw[dotted] (0.866,0.7) -- (1.039,0.8);
       \draw[dotted] (0.866,-0.7) -- (1.039,-0.8);
      \draw[dotted] (0.5,1.066) -- (0.6,1.239);
        \draw[dotted] (0.5,-1.066) -- (0.6,-1.239);
           \draw[dotted] (0,1.2) -- (0,1.4);
              \draw[dotted] (0,-1.2) -- (0,-1.4);
          \draw[dotted] (-0.5,1.066) -- (-0.6,1.239);
        \draw[dotted] (-0.5,-1.066) -- (-0.6,-1.239);
         \draw[dotted] (-0.866,0.7) -- (-1.039,0.8);
           \draw[dotted] (-0.866,-0.7) -- (-1.039,-0.8);
           \draw[dotted] (-1.4,0.2) -- (-1.4,0.4);
             \draw[dotted] (-1.4,-0.2) -- (-1.4,-0.4);
                     \draw[dotted] (-1.8,0.2) -- (-1.8,0.4);
             \draw[dotted] (-1.8,-0.2) -- (-1.8,-0.4);
              \draw[dotted] (-2.2,0.2) -- (-2.2,0.4);
             \draw[dotted] (-2.2,-0.2) -- (-2.2,-0.4);
      \draw (0.5,-0.8) node{$\chi$};
       \draw (0.7,-1.4) node{$\psi$};

\draw[dashed] (-2.5,0.2) arc (90:270:0.2 and 0.2);

\draw[solid] (1,0.2) arc (90:-90:0.01 and 0.2);

%insert fermions

\draw[red] (-1.9,0.3) -- (-1.7,0.1);
\draw[red] (-1.9,0.1) -- (-1.7,0.3);

\draw[red] (-1.9 ,0.3 -0.4) -- (-1.7 ,0.1 - 0.4);
\draw[red] (-1.9 ,0.1 - 0.4) -- (-1.7,0.3 -0.4);

%eqns

\draw (-3.8,-1.7) node{$\rho_{B,1} = $};

% divide

\draw (-3 , -1.7) -- (2,-1.7);

% normalization

\begin{scope}[yshift = -3.5cm]

\draw (-1.7,0.2) node{$\frac{1}{2} \times$};

 \draw (1,0.2) arc(0:360:1 and 1);

   \draw (1.2,0.2) arc(0:360:1.2 and 1.2);

      \draw[dotted] (1,0.2) -- (1.2,0.2);
      \draw[dotted] (-1,0.2) -- (-1.2,0.2);
     \draw[dotted] (0.866,0.7) -- (1.039,0.8);
       \draw[dotted] (0.866,-0.3) -- (1.039,-0.4);
      \draw[dotted] (0.5,1.066) -- (0.6,1.239);
        \draw[dotted] (0.5,-1.066 + 0.4) -- (0.6,-1.239 + 0.4);
           \draw[dotted] (0,1.2) -- (0,1.4);
              \draw[dotted] (0,-1.2 + 0.4) -- (0,-1.4 + 0.4);
          \draw[dotted] (-0.5,1.066) -- (-0.6,1.239);
        \draw[dotted] (-0.5,-1.066 + 0.4) -- (-0.6,-1.239 + 0.4);
         \draw[dotted] (-0.866,0.7) -- (-1.039,0.8);
           \draw[dotted] (-0.866,-0.7 + 0.4) -- (-1.039,-0.8 + 0.4);
      \draw (0.5,-0.8 + 0.4) node{$\chi$};
       \draw (0.7,-1.4 + 0.4) node{$\psi$};

\end{scope}
\end{tikzpicture}
}
\hspace{0.05\textwidth}
\subfigure[]  
{    
  \begin{tikzpicture}[thick,scale = 1]

 \draw (1,0.2) arc(0:180:1 and 1);
 \draw (1,-0.2) arc(0:-180:1 and 1);
  \draw (-1,-0.2) -- (-2.5,-0.2);
   \draw (-1,0.2) -- (-2.5,0.2);

   \draw (1.2,0.2) arc(0:170:1.2 and 1.2);
   \draw (1.2,-0.2) arc(0:-170:1.2 and 1.2);
   
    \draw (-1.18,-0.4) -- (-2.5,-0.4);
     \draw (-1.18,0.4) -- (-2.5,0.4);
     
     \draw[dotted] (0.866,0.7) -- (1.039,0.8);
       \draw[dotted] (0.866,-0.7) -- (1.039,-0.8);
      \draw[dotted] (0.5,1.066) -- (0.6,1.239);
        \draw[dotted] (0.5,-1.066) -- (0.6,-1.239);
           \draw[dotted] (0,1.2) -- (0,1.4);
              \draw[dotted] (0,-1.2) -- (0,-1.4);
          \draw[dotted] (-0.5,1.066) -- (-0.6,1.239);
        \draw[dotted] (-0.5,-1.066) -- (-0.6,-1.239);
         \draw[dotted] (-0.866,0.7) -- (-1.039,0.8);
           \draw[dotted] (-0.866,-0.7) -- (-1.039,-0.8);
           \draw[dotted] (-1.4,0.2) -- (-1.4,0.4);
             \draw[dotted] (-1.4,-0.2) -- (-1.4,-0.4);
                     \draw[dotted] (-1.8,0.2) -- (-1.8,0.4);
             \draw[dotted] (-1.8,-0.2) -- (-1.8,-0.4);
              \draw[dotted] (-2.2,0.2) -- (-2.2,0.4);
             \draw[dotted] (-2.2,-0.2) -- (-2.2,-0.4);
      \draw (0.5,-0.8) node{$\chi$};
       \draw (0.7,-1.4) node{$\psi$};

\draw[dashed] (-2.5,0.2) arc (90:270:0.2 and 0.2);

\draw[solid] (1,0.2) arc (90:-90:0.01 and 0.2);

%insert fermions

%eqns

\draw (-3.8,-1.7) node{$\rho_{B,2} = $};

% divide

\draw (-3 , -1.7) -- (2,-1.7);

% normalization

\begin{scope}[yshift = -3.5cm,xshift = -0.5 cm]

 \draw (1,0.2) arc(0:360:1 and 1);

   \draw (1.2,0.2) arc(0:360:1.2 and 1.2);

      \draw[dotted] (1,0.2) -- (1.2,0.2);
      \draw[dotted] (-1,0.2) -- (-1.2,0.2);
     \draw[dotted] (0.866,0.7) -- (1.039,0.8);
       \draw[dotted] (0.866,-0.3) -- (1.039,-0.4);
      \draw[dotted] (0.5,1.066) -- (0.6,1.239);
        \draw[dotted] (0.5,-1.066 + 0.4) -- (0.6,-1.239 + 0.4);
           \draw[dotted] (0,1.2) -- (0,1.4);
              \draw[dotted] (0,-1.2 + 0.4) -- (0,-1.4 + 0.4);
          \draw[dotted] (-0.5,1.066) -- (-0.6,1.239);
        \draw[dotted] (-0.5,-1.066 + 0.4) -- (-0.6,-1.239 + 0.4);
         \draw[dotted] (-0.866,0.7) -- (-1.039,0.8);
           \draw[dotted] (-0.866,-0.7 + 0.4) -- (-1.039,-0.8 + 0.4);
      \draw (0.5,-0.8 + 0.4) node{$\chi$};
       \draw (0.7,-1.4 + 0.4) node{$\psi$};

\end{scope}

\end{tikzpicture}
}

  \caption{We prepare two density matrices $\rho_{B,1}$ and $\rho_{B,2}$, where for $\rho_{B,1}$ we inserted a fermion operator $\chi_i$ at time $t_0$ (represented by the red crosses). In the figure, the denominators are the proper normalization factors, while the factor $1/2$ comes from $\chi^{2}_i = 1/2$.} 
  \label{twobath}   
 \end{figure}
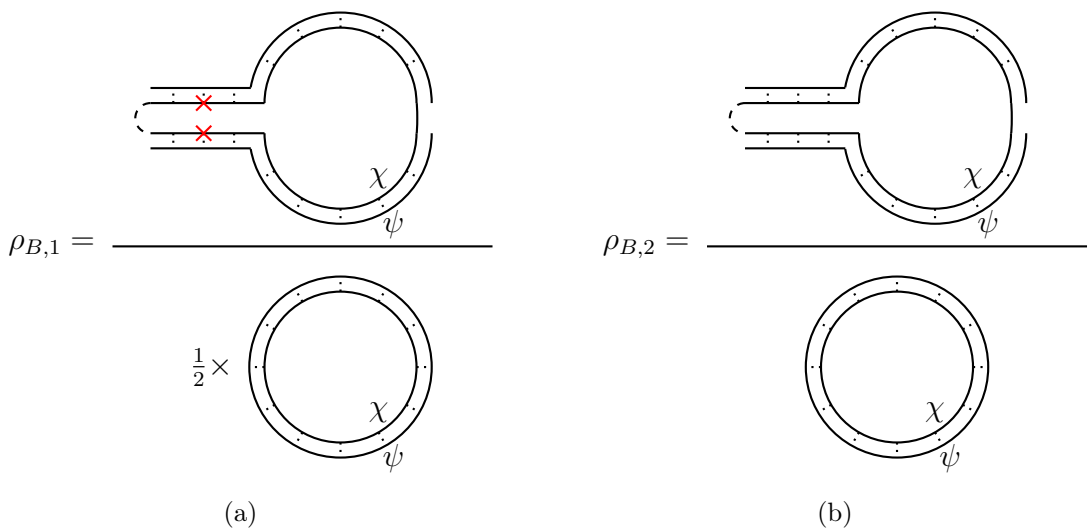

In the quantum mechanical system, throwing a particle into the black hole corresponds to adding a perturbation $\delta U$ to the evolution of the TFD. For simplicity, we choose the perturbation to be $\delta U=\sqrt{2}\chi_1$, applied at time $t_0$. We then study its effect on the state of the bath at a later time $t\geq t_0$. 
%The density matrix $\rho_{B,1}$ with the perturbation $\delta U$ is multiplied by a pair of additional Majorana operators at time $t_0$. 
The set up is illustrated in fig. \ref{twobath}. 

We denote the bath reduced density matrix at time $t$ in the perturbed case as $\rho_{B,1}(t)$, and the unperturbed one as $\rho_{B,2}(t)$. Obviously, $\rho_{B,1}(t_0)=\rho_{B,2}(t_0)$ since tracing over the system cancels $\delta U$ with $\delta U^\dagger$. For $t>t_0$ the effect is generically nontrivial. The informational retrieval depends on the distinguishibility of $\rho_{B,1}$ and $\rho_{B,2}$.

We would like to define an appropriate quantum information measure for the distinguishability of the two density operators. For this purpose, we introduce an ancilla qubit $A$ with two internal states $|0\rangle$ and $|1\rangle$. We initialize $A$ in a maximally mixed state $\rho_A=\frac12\mathbb{I}_2$, where $\mathbb{I}_2$ is the identity matrix with dimension $2$. We then perform a classically control operation: if the A system is in the state $|0\rangle$ ($|1\rangle$), we prepare a system without (with) perturbation $\delta U$, respectively. Tracing out the system $\chi$, the reduced density matrix for the ancilla qubit $A$ and the bath is given by
\begin{align}
\rho_{AB}(t)=\frac{1}{2}\begin{pmatrix}
\rho_{B,2}(t)&0\\
0&\rho_{B,1}(t)
\end{pmatrix}.
\end{align}
We can then compute the second R\'{e}nyi mutual information $I^{(2)}_{A,B}(t)$ between the bath and the control bit $A$, which quantifies whether we are able to reconstruct the perturbation by measuring the bath.\footnote{Other R\'{e}nyi mutual information and von Neumann mutual information quantities can also be studied, but we will focus on the second R\'{e}nyi mutual information since it can be computed directly from our two-replica numerics.} The relevant entropy quantities are given by:
\begin{align}
&\exp\left(-S^{(2)}_B(t)\right)=\frac{1}{4}\text{tr}\left(\rho_{B,1}^2(t)+\rho_{B,2}^2(t)+2\rho_{B,1}(t)\rho_{B,2}(t)\right), \\
&\exp\left(-S^{(2)}_A(t)\right)=\frac{1}{2},\ \ \ \ \exp\left(-S^{(2)}_{AB}(t)\right)=\frac{1}{4}\text{tr}\left(\rho_{B,1}^2(t)+\rho_{B,2}^2(t)\right).
\end{align}
As a result, we have
\begin{align}
I^{(2)}_{A,B}(t)=-\log\left(\frac{1}{2}+\frac{\text{tr}\left(\rho_{B,1}(t)\rho_{B,2}(t)\right)}{\text{tr}\left(\rho_{B,1}(t)^2\right)+\text{tr}\left(\rho_{B,2}(t)^2\right)}\right). \label{mutual}
\end{align}
Right after the insertion of the Majorana operator, we have $\rho_{B,1}(t_0)=\rho_{B,2}(t_0)$ and $I^{(2)}_{A,B}=0$. As a result, the bath is not affected by the perturbation. On the contrary, from the gravity picture, we expect that in the long time limit, density matrix with and without the perturbation becomes orthogonal $\text{tr}\left(\rho_{B,1}(t)\rho_{B,2}(t)\right)=0$, which leads to $I^{(2)}_{A,B}=\log(2)$.

To study \eqref{mutual}, we could further express both ${\text{tr}\left(\rho_{B,1}(t)\rho_{B,2}(t)\right)}/{\text{tr}\left(\rho_{B,2}(t)^2\right)}$ and ${\text{tr}\left(\rho_{B,1}(t)^2\right)}/{\text{tr}\left(\rho_{B,2}(t)^2\right)}$ in terms of the Green's functions of Majorana fermions on the contour in fig. \ref{fig:fig2}. As an example (see fig. \ref{twobath2}), we have:

\begin{figure}[t]
  \center

    \begin{tikzpicture}[thick,scale = 0.65]

\node[] at (-5.5,1.5) {  $\frac{\textrm{tr}(\rho_{B,1}\rho_{B,2})}{2\textrm{tr}(\rho_{B,2}^2)} = $};
 
\node[] at (12,1.5) {  $= G(2\beta + 7t +t_0 , t-t_0)$}; 
 
  \draw (1,0.2) arc(0:180:1 and 1);
 \draw (1,-0.2) arc(0:-180:1 and 1);
  \draw (-1,-0.2) -- (-2.5,-0.2);
   \draw (-1,0.2) -- (-2.5,0.2);

   \draw (1.2,0.2) arc(0:170:1.2 and 1.2);
   \draw (1.2,-0.2) arc(0:-170:1.2 and 1.2);
 
    \draw (-1.18,-0.4) -- (-2.5,-0.4);
     \draw (-1.18,0.4) -- (-2.5,0.4);
  
     \draw[dotted] (0.866,0.7) -- (1.039,0.8);
       \draw[dotted] (0.866,-0.7) -- (1.039,-0.8);
      \draw[dotted] (0.5,1.066) -- (0.6,1.239);
        \draw[dotted] (0.5,-1.066) -- (0.6,-1.239);
           \draw[dotted] (0,1.2) -- (0,1.4);
              \draw[dotted] (0,-1.2) -- (0,-1.4);
          \draw[dotted] (-0.5,1.066) -- (-0.6,1.239);
        \draw[dotted] (-0.5,-1.066) -- (-0.6,-1.239);
         \draw[dotted] (-0.866,0.7) -- (-1.039,0.8);
           \draw[dotted] (-0.866,-0.7) -- (-1.039,-0.8);
           \draw[dotted] (-1.4,0.2) -- (-1.4,0.4);
             \draw[dotted] (-1.4,-0.2) -- (-1.4,-0.4);
                     \draw[dotted] (-1.8,0.2) -- (-1.8,0.4);
             \draw[dotted] (-1.8,-0.2) -- (-1.8,-0.4);
              \draw[dotted] (-2.2,0.2) -- (-2.2,0.4);
             \draw[dotted] (-2.2,-0.2) -- (-2.2,-0.4);
      \draw (0.5,-0.8) node{$\chi$};
       \draw (0.7,-1.4) node{$\psi$};

  %boundary conditions
 \draw[dashed] (-2.5,0.2) arc (90:270:0.2 and 0.2);

\draw[solid] (1,0.2) arc (90:-90:0.01 and 0.2);

 \begin{scope}[shift={(0,3)}]
 \draw (1,0.2) arc(0:180:1 and 1);
 \draw (1,-0.2) arc(0:-180:1 and 1);
  \draw (-1,-0.2) -- (-2.5,-0.2);
   \draw (-1,0.2) -- (-2.5,0.2);

   \draw (1.2,0.2) arc(0:170:1.2 and 1.2);
   \draw (1.2,-0.2) arc(0:-170:1.2 and 1.2);
   
    \draw (-1.18,-0.4) -- (-2.5,-0.4);
     \draw (-1.18,0.4) -- (-2.5,0.4);

     \draw[dotted] (0.866,0.7) -- (1.039,0.8);
       \draw[dotted] (0.866,-0.7) -- (1.039,-0.8);
      \draw[dotted] (0.5,1.066) -- (0.6,1.239);
        \draw[dotted] (0.5,-1.066) -- (0.6,-1.239);
           \draw[dotted] (0,1.2) -- (0,1.4);
              \draw[dotted] (0,-1.2) -- (0,-1.4);
          \draw[dotted] (-0.5,1.066) -- (-0.6,1.239);
        \draw[dotted] (-0.5,-1.066) -- (-0.6,-1.239);
         \draw[dotted] (-0.866,0.7) -- (-1.039,0.8);
           \draw[dotted] (-0.866,-0.7) -- (-1.039,-0.8);
           \draw[dotted] (-1.4,0.2) -- (-1.4,0.4);
             \draw[dotted] (-1.4,-0.2) -- (-1.4,-0.4);
                     \draw[dotted] (-1.8,0.2) -- (-1.8,0.4);
             \draw[dotted] (-1.8,-0.2) -- (-1.8,-0.4);
              \draw[dotted] (-2.2,0.2) -- (-2.2,0.4);
             \draw[dotted] (-2.2,-0.2) -- (-2.2,-0.4);
      \draw (0.5,-0.8) node{$\chi$};
       \draw (0.7,-1.4) node{$\psi$};
 
 %fermion
  
  \draw[red] (-1.9,0.3) -- (-1.7,0.1);
\draw[red] (-1.9,0.1) -- (-1.7,0.3);

\draw[red] (-1.9 ,0.3 -0.4) -- (-1.7 ,0.1 - 0.4);
\draw[red] (-1.9 ,0.1 - 0.4) -- (-1.7,0.3 -0.4);
  
 %boundary conditions
 \draw[dashed] (-2.5,0.2) arc (90:270:0.2 and 0.2);

\draw[solid] (1,0.2) arc (90:-90:0.01 and 0.2);

\end{scope}

%boundary conditions up and down

\draw[dashed] (-2.5,0.4+3) arc(90:270:0.7 and 1.9);
\draw[dashed] (-2.5,-0.4+3) arc(90:270:0.5 and 1.1);

\draw[dashed] (1.2,0.2+3) arc(90:-90:0.7 and 1.7);
\draw[dashed] (1.2, -0.2+3) arc(90:-90:0.5 and 1.3);

%divide

\draw (2,-1.5) -- (3,4.5);

%normalization
\begin{scope}[shift={(6.4,0)}]
 
  \draw (1,0.2) arc(0:180:1 and 1);
 \draw (1,-0.2) arc(0:-180:1 and 1);
  \draw (-1,-0.2) -- (-2.5,-0.2);
   \draw (-1,0.2) -- (-2.5,0.2);

   \draw (1.2,0.2) arc(0:170:1.2 and 1.2);
   \draw (1.2,-0.2) arc(0:-170:1.2 and 1.2);
 
    \draw (-1.18,-0.4) -- (-2.5,-0.4);
     \draw (-1.18,0.4) -- (-2.5,0.4);
  
     \draw[dotted] (0.866,0.7) -- (1.039,0.8);
       \draw[dotted] (0.866,-0.7) -- (1.039,-0.8);
      \draw[dotted] (0.5,1.066) -- (0.6,1.239);
        \draw[dotted] (0.5,-1.066) -- (0.6,-1.239);
           \draw[dotted] (0,1.2) -- (0,1.4);
              \draw[dotted] (0,-1.2) -- (0,-1.4);
          \draw[dotted] (-0.5,1.066) -- (-0.6,1.239);
        \draw[dotted] (-0.5,-1.066) -- (-0.6,-1.239);
         \draw[dotted] (-0.866,0.7) -- (-1.039,0.8);
           \draw[dotted] (-0.866,-0.7) -- (-1.039,-0.8);
           \draw[dotted] (-1.4,0.2) -- (-1.4,0.4);
             \draw[dotted] (-1.4,-0.2) -- (-1.4,-0.4);
                     \draw[dotted] (-1.8,0.2) -- (-1.8,0.4);
             \draw[dotted] (-1.8,-0.2) -- (-1.8,-0.4);
              \draw[dotted] (-2.2,0.2) -- (-2.2,0.4);
             \draw[dotted] (-2.2,-0.2) -- (-2.2,-0.4);
      \draw (0.5,-0.8) node{$\chi$};
       \draw (0.7,-1.4) node{$\psi$};

  %boundary conditions
 \draw[dashed] (-2.5,0.2) arc (90:270:0.2 and 0.2);

\draw[solid] (1,0.2) arc (90:-90:0.01 and 0.2);

 \begin{scope}[shift={(0,3)}]
 \draw (1,0.2) arc(0:180:1 and 1);
 \draw (1,-0.2) arc(0:-180:1 and 1);
  \draw (-1,-0.2) -- (-2.5,-0.2);
   \draw (-1,0.2) -- (-2.5,0.2);

   \draw (1.2,0.2) arc(0:170:1.2 and 1.2);
   \draw (1.2,-0.2) arc(0:-170:1.2 and 1.2);
   
    \draw (-1.18,-0.4) -- (-2.5,-0.4);
     \draw (-1.18,0.4) -- (-2.5,0.4);

     \draw[dotted] (0.866,0.7) -- (1.039,0.8);
       \draw[dotted] (0.866,-0.7) -- (1.039,-0.8);
      \draw[dotted] (0.5,1.066) -- (0.6,1.239);
        \draw[dotted] (0.5,-1.066) -- (0.6,-1.239);
           \draw[dotted] (0,1.2) -- (0,1.4);
              \draw[dotted] (0,-1.2) -- (0,-1.4);
          \draw[dotted] (-0.5,1.066) -- (-0.6,1.239);
        \draw[dotted] (-0.5,-1.066) -- (-0.6,-1.239);
         \draw[dotted] (-0.866,0.7) -- (-1.039,0.8);
           \draw[dotted] (-0.866,-0.7) -- (-1.039,-0.8);
           \draw[dotted] (-1.4,0.2) -- (-1.4,0.4);
             \draw[dotted] (-1.4,-0.2) -- (-1.4,-0.4);
                     \draw[dotted] (-1.8,0.2) -- (-1.8,0.4);
             \draw[dotted] (-1.8,-0.2) -- (-1.8,-0.4);
              \draw[dotted] (-2.2,0.2) -- (-2.2,0.4);
             \draw[dotted] (-2.2,-0.2) -- (-2.2,-0.4);
      \draw (0.5,-0.8) node{$\chi$};
       \draw (0.7,-1.4) node{$\psi$};

 %boundary conditions
 \draw[dashed] (-2.5,0.2) arc (90:270:0.2 and 0.2);

\draw[solid] (1,0.2) arc (90:-90:0.01 and 0.2);

\end{scope}

\draw[dashed] (-2.5,0.4+3) arc(90:270:0.7 and 1.9);
\draw[dashed] (-2.5,-0.4+3) arc(90:270:0.5 and 1.1);

\draw[dashed] (1.2,0.2+3) arc(90:-90:0.7 and 1.7);
\draw[dashed] (1.2, -0.2+3) arc(90:-90:0.5 and 1.3);

\end{scope}

\end{tikzpicture}

  \caption{We illustrate how the overlap $\textrm{tr} (\rho_{B,1}(t)\rho_{B,2}(t))$ is related to our saddle point solution $G$, to the leading order in $N$. } 
  \label{twobath2}  
 \end{figure}

\begin{align}\label{overlapgreen}
\frac{\text{tr}\left(\rho_{B,1}(t)\rho_{B,2}(t)\right)}{\text{tr}\left(\rho_{B,2}(t)^2\right)}=2\left<\chi_1(2\beta+7t+t_0)\chi_1(t-t_0)\right>=2G(2\beta+7t+t_0,t-t_0).
\end{align}
Note that, although the mutual information is more physical, this overlap, which is equivalent to the Green's function, also measures the difference between the density matrices.

Similarly, $\text{tr}\left(\rho_{B,1}(t)^2\right)/\text{tr}\left(\rho_{B,2}(t)^2\right)$ can be expressed as a four-point function, the leading order contribution of which is given by the factorized result:
\begin{align}
\frac{\text{tr}\left(\rho_{B,1}(t)^2\right)}{\text{tr}\left(\rho_{B,2}(t)^2\right)}=&4\left<\chi_1(2\beta+7t+t_0)\chi_1(\beta+5t-t_0)\chi_1(\beta+3t+t_0)\chi_1(t-t_0)\right>\notag\\
=&4\left[ G(2\beta+7t+t_0,t-t_0)G(\beta+5t-t_0,\beta+3t+t_0) \right. \notag\\
&-G(2\beta+7t+t_0,\beta+3t+t_0)G(\beta+5t-t_0,t-t_0) \notag\\
& \left. + G(2\beta+7t+t_0,\beta+5t-t_0)G(\beta+3t+t_0,t-t_0)\right]. 
\end{align}

\begin{figure}[t]
  \center
  \includegraphics[width=1\columnwidth]{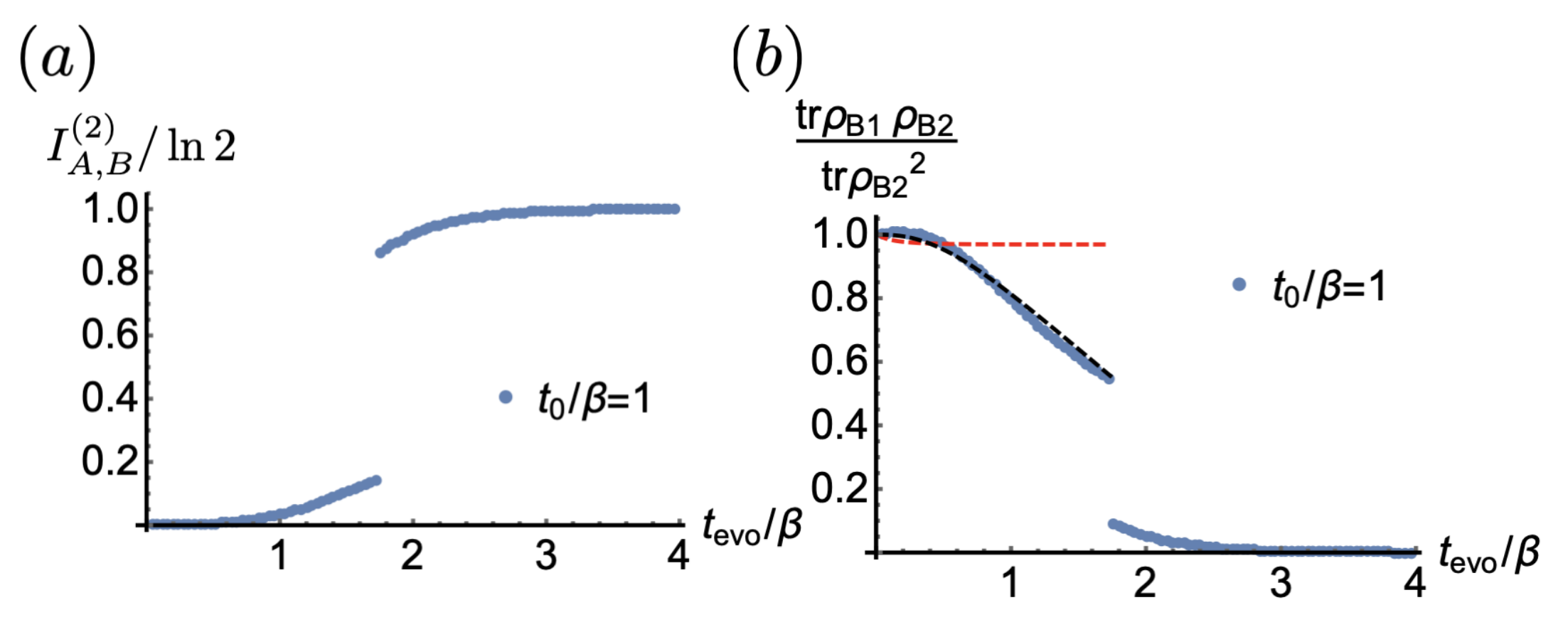}
  \caption{Numerical results on the R\'{e}nyi mutual information and the overlap for $t_0=\beta<t_{\text{Page}}$ (blue dots), with $V^2/J = 0.05, \beta J=4, \Lambda=5J$. The black dashed line is the analytical result using the Schwarzian four point function and the red dashed line is the result coming from the disconnected four point function.} 
  \label{overlap}   
 \end{figure}

From the discussion in the previous sections, we learned that across the phase transition at the Page time, the Green's function changes discontinuously. In fig. \ref{overlap}, we show an example of how the the mutual information and the overlap looks like if we add the pertutbation at $t_0=\beta$. It should be noted that the R\'{e}nyi mutual information is nonzero even before the Page time, which is because the particle created by $\chi_1(t_0)$ has a finite probability to directly hop to the bath. In other words, $\chi_1(t_0)$ creates a particle that is not entirely infalling, but has a finite probability of going out into the bath. The contribution of this direct coupling to the R\'{e}nyi mutual information is proportional to $V^2$. 

\begin{figure}[t]
  \center
  \includegraphics[width=1\columnwidth]{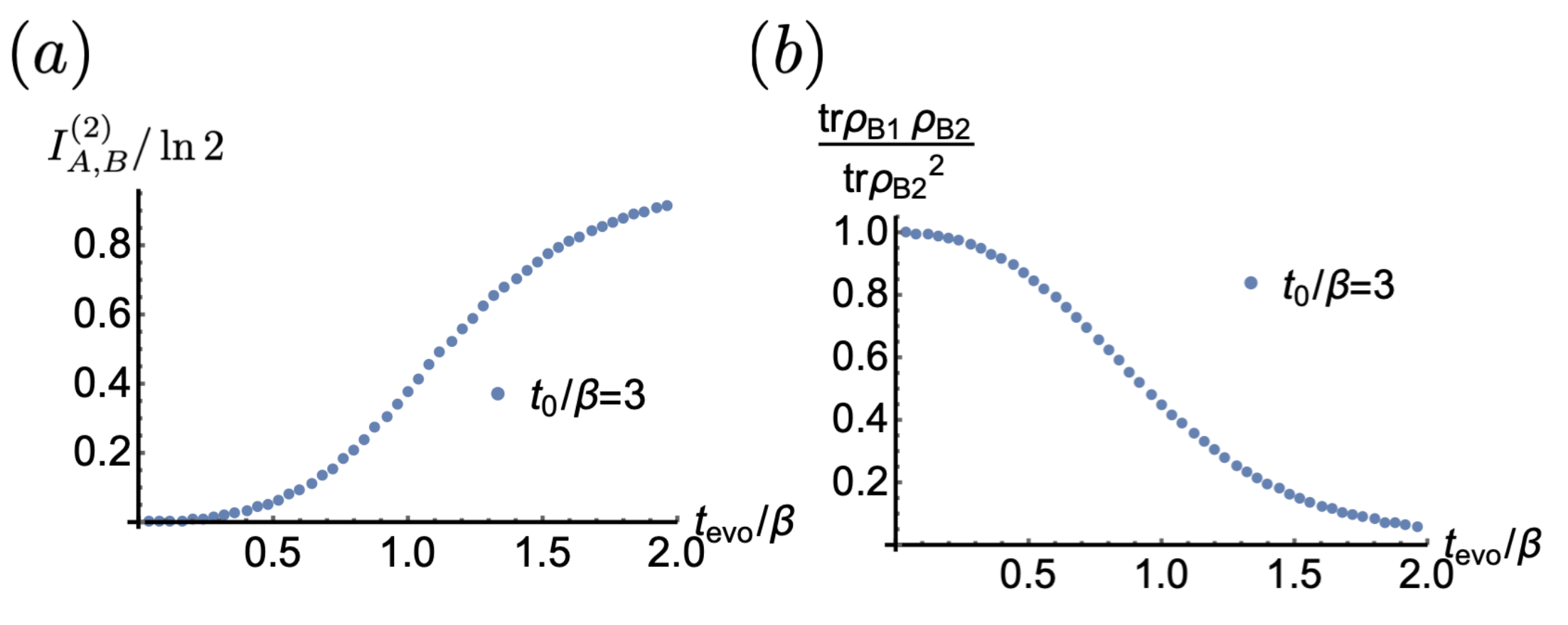}
  \caption{Numerical results of the mutual information and the overlap (blue dots), with $V^2/J = 0.05, \beta J=4, \Lambda=5J$ for a late perturbation at $t_0=3\beta>t_{\text{Page}}$.} 
  \label{overlap2}   
 \end{figure}

At the Page time, both the mutual information and the overlap are discontinuous. The jumps correspond to the fact that after the Page time, the information carried by the infalling particle is encoded in the bath in a nonlocal way. If we throw in a particle before the Page time, most information will only be retrievable after Page time. In comparison, if we throw in the information after the transition, we only need to wait for a scrambling time $t_{sc}\sim \beta$, as shown in fig. \ref{overlap2}. Note that because the bath has a central charge $N$, the information retrieval time is order $1$ rather than order $\log N$. In the bulk picture \cite{almheiri2019islands}, this corresponds to an island that is finite distance outside the horizon. 

In Fig. \ref{overlap}, we have also plotted the perturbative result for the Green’s function in (\ref{overlapgreen}). Focusing on a single replica, the change of the Green’s function comes from the effective action \eqref{action short}. We copy it here for convenience:
\begin{align}
\delta S=NV^2 \int_{\mathcal{C}_1}d\tau_1\int_{\mathcal{C}_2}d\tau_2\ G(\tau_1,\tau_2) g(\tau_1,\tau_2).
\end{align}
 To the $V^2/J$ order, we have the change of the Green's function being:
\begin{align}
    \delta G(\tau_1,\tau_2)&=-\left<G(\tau_1,\tau_2)\delta S\right>+\left<G(\tau_1,\tau_2)\right>\left<\delta S\right>\notag \\&=-\left<G(\tau_1,\tau_2)\delta S\right>_c\notag \\&=-NV^2 \int_{\mathcal{C}_1}d\tau_3\int_{\mathcal{C}_2}d\tau_4 \left<G(\tau_1,\tau_2)G(\tau_3,\tau_4)\right>_c g(\tau_3,\tau_4).
\end{align}
Note that the connected part of four-point function is $1/N$, which cancels the factor $N$ in front. Instead of using the imaginary time and applying an analytical continuation at the end, here we directly include the real-time contour in the calculation. We set $\tau_1=\beta/2+\epsilon/2+i t_0$, $\tau_2=\beta/2-\epsilon/2+i t_0$, $\tau_3=\beta/2+\epsilon/2+i t_3$ and $\tau_4=\beta/2-\epsilon/2+i t_4$.  The main time dependence is from:

\begin{equation}\label{deltaG}
\begin{tikzpicture}[thick,scale = 0.8,baseline={([yshift=-3pt]current bounding box.center)}]
\draw (-0.3+1,-0.2) arc (-90:90:0.1 and 0.2);

\draw (-0.3+1,0.2) --(-0.3-0.1-1.2,0.2) ;
\draw  (-0.3+1,-0.2) --(-0.3-0.1-1.2,-0.2) ;

\draw[thick,->,>=stealth]  (-0.3-0.1-1.2,0.2) --(-0.3-0.15,0.2) ;
\draw[thick,->,>=stealth]  (-0.3+1,-0.2) --(-0.3-0.15,-0.2) ;

\draw  (-0.3-0.1-1.2,0.16)  arc (11:360-11:1 and 1);
\node at (-1,-0.2)[circle,fill,inner sep=1.5pt]{};
\node at (-1,0.2)[circle,fill,inner sep=1.5pt]{};
\node at (0,-0.2)[circle,fill,inner sep=1.5pt]{};
\node at (0,0.2)[circle,fill,inner sep=1.5pt]{};

\draw (-1,-0.6) node{$2$}; 
\draw (-1,0.6) node{$1$}; 

\draw (0,-0.6) node{$4$}; 
\draw (0,0.6) node{$3$};
\draw (1,0) node{$t$};
\end{tikzpicture}=-\frac{V^2}{\pi}\int_{t_0}^{t}dt_3\int_{t_0}^{t}dt_4F_c(\tau_1,\tau_2,\tau_3,\tau_4)\frac{\pi}{\beta \sin\frac{\pi(\tau_3-\tau_4)}{\beta}} .
\end{equation}
Here we use the four-point function $F_c$ from the Schwarizan theory (for $\beta=2\pi$) \cite{maldacena2016remarks,kitaev2018soft}:
\begin{align}
    F_c(\theta_1,\theta_2,\theta_3,\theta_4)=&\frac{J}{2 \sqrt{2} \pi  \alpha _S}b^2\times\notag \\&\frac{\left(2 \sin \left(\frac{\theta _{12}}{2}\right)+\left(2 \pi -\theta _{12}\right)
   \cos \left(\frac{\theta _{12}}{2}\right)\right) \left(2 \sin \left(\frac{\theta
   _{34}}{2}\right)-\theta _{34} \cos \left(\frac{\theta _{34}}{2}\right)\right)}{32
   \sin ^{\frac{3}{2}}\left(\frac{\theta _{12}}{2}\right) \sin
   ^{\frac{3}{2}}\left(\frac{\theta _{34}}{2}\right)}.
\end{align}
We evaluate the integral in eq. \eqref{deltaG} numerically since there is no closed analytic form. We choose the cutoff $\epsilon=\frac{ c_0}{2\pi J}$ and adjust $c_0$ to match with the numerical result. For $c_0=4$, the result is shown in fig. \ref{overlap} (a) as the black dashed line, which suggests that the estimation works reasonably well. In the long time limit $t\gg \beta$, the first order perturbative result is linear in time:
\begin{align}
    \delta G/G\sim -\frac{b \sqrt{\beta } J t V^2 \Gamma \left(\frac{7}{4}\right)}{18 \sqrt{2} \pi 
   \epsilon_\theta  \Gamma \left(\frac{5}{4}\right) \alpha _S}.
\end{align}
with $\epsilon_\theta=\epsilon\frac{2\pi}{\beta}=\frac{c_0}{\beta J}$.

 The linear decrease comes from the fact that for $\delta{G}(\tau_1,\tau_2)$ with $\tau_1=\beta/2+\epsilon/2+i t_0$, $\tau_2=\beta/2-\epsilon/2+i t_0$, the four-point function is finite even if $t_3\sim t_4 \gg t_0$. Physically, the reason is that the infalling particle carries an $SL(2,R)$ momentum, and the boundary will gain an opposite momentum, required by the overall $SL(2,R)$ symmetry. Consequently, the backreaction induced by the infalling particle does not decay with time.

\begin{figure}
    \centering
    \includegraphics[width=5.5in]{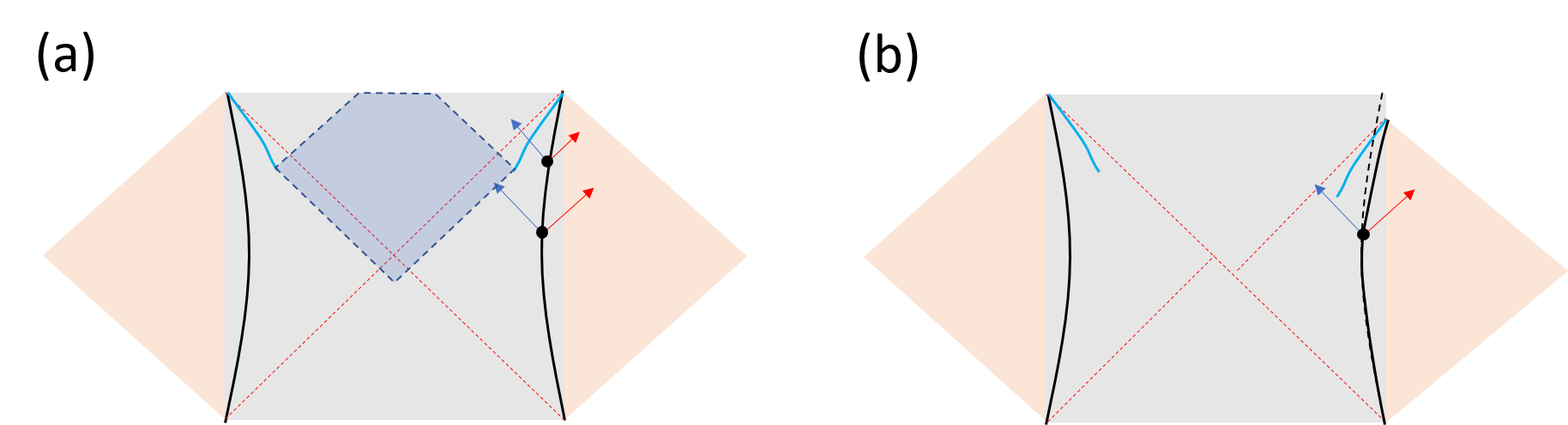}
    \caption{Bulk dual interpretation of the boundary perturbation of adding a $\chi_i$ fermion. We sketch the Penrose diagram of an AdS$_2$ eternal black hole coupled with the flat space bath, discussed in Ref. \cite{almheiri2019islands}. The blue solid curve indicates the worldline of the boundary of the island. (a) When backreaction is neglected, this perturbation creates a superposition of infalling fermion mode (blue arrow) and outgoing fermion mode (red arrow). The blue diamond is contained in the entanglement wedge of the bath at the Page time. A particle created earlier than the Page time can be retrieved from the bath soon after the Page time, while a particle created after Page time can be retrieved after a finite scrambling time. (b) Considering the backreaction, the infalling fermion induces a change of the boundary location, described by the Schwarzian action in low energy. This leads to a decreasing overlap between the perturbed and unperturbed states of the bath even before the Page time (see text).}
    \label{fig:bulkillustration}
\end{figure}

For comparison, we also study the result if we approximate the four point function by only the disconnected part:
\begin{align}
F_c(\theta_1,\theta_2,\theta_3,\theta_4)=G(\theta_1,\theta_3)G(\theta_4,\theta_2)-G(\theta_1,\theta_4)G(\theta_3,\theta_2).
\end{align}
In the bulk interpretation, this corresponds to neglecting the backreaction and considering a free fermion problem. In this case, it is easy to see that $F_c \rightarrow 0$ in the limit $t_3\sim t_4 \gg t_0$. The free fermion result is also plotted in Fig. \ref{overlap} (b) by the red dashed line. Instead of linear $t$ dependence, the overlap saturates to a finite value (until the Page time), corresponding to a finite probability of the initial particle moving outwards. Comparing the two approximations, we see that the change of the overlap before Page time is mainly due to the backreaction.

Comparing the R\'{e}nyi mutual information and the overlap, we see that the mutual information changes much slower for short time. This is because in the short time limit, $V^2/J$ contribution for the mutual information vanishes:
\begin{align}
I^{(2)}_{A,B}(t)\approx-\log\left(\frac{1}{2}+\frac{1+2\delta G}{1+(1+2\delta G)^2}+\mathcal{O}(V^4/J^2)\right)= \mathcal{O}(V^4/J^2). 
\end{align}
The decrease of $I^{(2)}_{A,B}$ then comes from higher order corrections including correlation between different contours.

\section{Conclusion and Discussion}\label{sec:conclusion}

In conclusion, we have studied the SYK model coupled to a free Majorana fermion bath, as a toy model to investigate the physics of black hole evaporation. We studied the time evolution of the two-point function and the second R\'{e}nyi entropy of the thermofield double state of this coupled system. For low coupling with the bath, we found a first order transition in the second R\'{e}nyi entropy, which corresponds to the formation of a "replica wormhole", similar to the results in the section 5 of Ref. \cite{penington2019replica}. We also studied the information retrieval from the black hole by creating a single fermion on the boundary. By comparing the perturbed and unperturbed reduced density operators of the thermal bath, we see a combination of two kinds of effects. Before the Page time, the bath already knows partially about the perturbation to the black hole system, because the boundary fermion has a finite chance to directly leak to the bath, and also because the backreaction of the infalling fermion. The latter effect makes dominant contribution. At the Page time, the information available to the bath about the perturbation has a finite jump, which is consistent with the expectation that the black hole almost saturates to its maximal entropy state after the Page time and therefore almost cannot preserve any information about the perturbation. 

With this concrete model, there are many open questions. Although we've shown that the information is in principle retrievable through quantum information argument, we've not provided an explicit construction of the form of the bulk fermion operator. It will be interesting to study the bulk fermion operators more explicitly, especially the fermions behind the horizon. If the bulk fermion operators can be identified, it may become possible to more explicitly investigate the black hole information paradox such as the firewall paradox \cite{almheiri2013black}. It would also be nice if one can see explicitly how the proposals for recovering operators in the island \cite{penington2019replica,Chen:2019iro} work in this set-up. Another question about the information retrieval is whether the backreaction effect we observed for two-replica calculation should vanish if we take the von Neumann limit. For example if we compute the relative entropy $S(\rho_{B1}|\rho_{B2})$, will the backreaction effect still be significant?

\paragraph{Acknowledgement}

We thank Pouria Dadras, Yingfei Gu, Alexei Kitaev, Juan Maldacena,  Stephen Shenker, Douglas Stanford, Zhenbin Yang, Shunyu Yao and Hui Zhai for helpful discussions. We thank Arjun Nachiket for suggestions on the manuscript. PZ would like to thank Pouria Dadras and Alexei Kitaev for discussions on their ongoing work, which focus on the von Neumann entropy dynamics of the coupled SYK models. This work is supported by the National Science Foundation Grant No. 1720504 and the Simons Foundation. This work is also supported in part by the DOE Office of Science, Office of High Energy Physics, the grant DE-SC0019380.

\begin{appendices}

\section{The Green's function of the bath on the Keldysh contour}\label{A}

In this appendix, we list the detailed expression for $g(s,s')$ depending on the different locations of $s$ and $s'$. Since $g(s,s')$ is only non zero when $s$ and $s'$ is on the same contour in fig.\ref{fig:fig2}(b), here we only draw a single Keldysh contour. We have:
\begin{equation}
\begin{tikzpicture}[thick,scale = 0.8,baseline={([yshift=-3pt]current bounding box.center)}]
\draw (-0.3+1,-0.2) arc (-90:90:0.1 and 0.2);

\draw (-0.3+1,0.2) --(-0.3-0.1-1.2,0.2) ;
\draw  (-0.3+1,-0.2) --(-0.3-0.1-1.2,-0.2) ;

\draw[thick,->,>=stealth]  (-0.3-0.1-1.2,0.2) --(-0.3-0.15,0.2) ;
\draw[thick,->,>=stealth]  (-0.3+1,-0.2) --(-0.3-0.15,-0.2) ;

\node at (-1,-0.2)[circle,fill,inner sep=1.5pt]{};
\node at (0.2,-0.2)[circle,fill,inner sep=1.5pt]{};
\draw  (-0.3-0.1-1.2,0.16)  arc (11:360-11:1 and 1);
\draw (-3,0) node{$\psi$}; 
\draw (-1,-0.6) node{$t_1$}; 
\draw (0.2,-0.6) node{$t_2$};
\end{tikzpicture}\ =\int^{\pi\Lambda}_{-\pi\Lambda}\frac{dk}{2\pi}\frac{e^{-i\epsilon_k(t_1-t_2)}}{1+e^{-\beta\epsilon_k}}\equiv g(t_1-t_2).\ \ \ \ \ \ \ \ \ \ 
\end{equation}
\begin{equation}
\begin{tikzpicture}[thick,scale = 0.8,baseline={([yshift=-3pt]current bounding box.center)}]
\draw (-0.3+1,-0.2) arc (-90:90:0.1 and 0.2);

\draw (-0.3+1,0.2) --(-0.3-0.1-1.2,0.2) ;
\draw  (-0.3+1,-0.2) --(-0.3-0.1-1.2,-0.2) ;

\draw[thick,->,>=stealth]  (-0.3-0.1-1.2,0.2) --(-0.3-0.15,0.2) ;
\draw[thick,->,>=stealth]  (-0.3+1,-0.2) --(-0.3-0.15,-0.2) ;

\draw  (-0.3-0.1-1.2,0.16)  arc (11:360-11:1 and 1);
\node at (-1.6,-0.2)[circle,fill,inner sep=1.5pt]{};
\node at (0.2,-0.2)[circle,fill,inner sep=1.5pt]{};
\node at (-3,-0.95)[circle,fill,inner sep=1.5pt]{};

\draw (-3,0) node{$\psi$}; 
\draw (-1.4,-0.6) node{$0$}; 
\draw (0.2,-0.6) node{$t_2$};
\draw (-2.6,-1.3) node{$\tau_1$};
\end{tikzpicture}\ =\int^{\pi\Lambda}_{-\pi\Lambda}\frac{dk}{2\pi}\frac{e^{-i\epsilon_k(-i\tau_1-t_2)}}{1+e^{-\beta\epsilon_k}}= g(-i\tau_1-t_2).\ \ \ \ 
\end{equation}
\begin{equation}
\begin{tikzpicture}[thick,scale = 0.8,baseline={([yshift=-3pt]current bounding box.center)}]
\draw (-0.3+1,-0.2) arc (-90:90:0.1 and 0.2);

\draw (-0.3+1,0.2) --(-0.3-0.1-1.2,0.2) ;
\draw  (-0.3+1,-0.2) --(-0.3-0.1-1.2,-0.2) ;

\draw[thick,->,>=stealth]  (-0.3-0.1-1.2,0.2) --(-0.3-0.15,0.2) ;
\draw[thick,->,>=stealth]  (-0.3+1,-0.2) --(-0.3-0.15,-0.2) ;

\node at (-1,0.2)[circle,fill,inner sep=1.5pt]{};
\node at (0.2,-0.2)[circle,fill,inner sep=1.5pt]{};
\draw  (-0.3-0.1-1.2,0.16)  arc (11:360-11:1 and 1);
\draw (-3,0) node{$\psi$}; 
\draw (-1,0.6) node{$t_1$}; 
\draw (0.2,-0.6) node{$t_2$};
\end{tikzpicture}\ =\int^{\pi\Lambda}_{-\pi\Lambda}\frac{dk}{2\pi}\frac{e^{-i\epsilon_k(t_2-t_1)}}{1+e^{-\beta\epsilon_k}}= g(t_2-t_1).\ \ \ \ \ \ \ \ \ \ 
\end{equation}
\begin{equation}
\begin{tikzpicture}[thick,scale = 0.8,baseline={([yshift=-3pt]current bounding box.center)}]
\draw (-0.3+1,-0.2) arc (-90:90:0.1 and 0.2);

\draw (-0.3+1,0.2) --(-0.3-0.1-1.2,0.2) ;
\draw  (-0.3+1,-0.2) --(-0.3-0.1-1.2,-0.2) ;

\draw[thick,->,>=stealth]  (-0.3-0.1-1.2,0.2) --(-0.3-0.15,0.2) ;
\draw[thick,->,>=stealth]  (-0.3+1,-0.2) --(-0.3-0.15,-0.2) ;

\draw  (-0.3-0.1-1.2,0.16)  arc (11:360-11:1 and 1);
\node at (-1.6,-0.2)[circle,fill,inner sep=1.5pt]{};
\node at (-3,0.9)[circle,fill,inner sep=1.5pt]{};
\node at (-3,-0.95)[circle,fill,inner sep=1.5pt]{};

\draw (-3,0) node{$\psi$}; 
\draw (-1.4,-0.6) node{$0$}; 
\draw (-2.6,1.2) node{$\tau_1$};
\draw (-2.6,-1.3) node{$\tau_2$};
\end{tikzpicture}\ =\int^{\pi\Lambda}_{-\pi\Lambda}\frac{dk}{2\pi}\frac{e^{-\epsilon_k(\tau_1-\tau_2)}}{1+e^{-\beta\epsilon_k}}= g(-i(\tau_1-\tau_2)).\ \ \ \ 
\end{equation}
\begin{equation}
\begin{tikzpicture}[thick,scale = 0.8,baseline={([yshift=-3pt]current bounding box.center)}]
\draw (-0.3+1,-0.2) arc (-90:90:0.1 and 0.2);

\draw (-0.3+1,0.2) --(-0.3-0.1-1.2,0.2) ;
\draw  (-0.3+1,-0.2) --(-0.3-0.1-1.2,-0.2) ;

\draw[thick,->,>=stealth]  (-0.3-0.1-1.2,0.2) --(-0.3-0.15,0.2) ;
\draw[thick,->,>=stealth]  (-0.3+1,-0.2) --(-0.3-0.15,-0.2) ;

\draw  (-0.3-0.1-1.2,0.16)  arc (11:360-11:1 and 1);
\node at (-1.6,-0.2)[circle,fill,inner sep=1.5pt]{};
\node at (-3,0.9)[circle,fill,inner sep=1.5pt]{};
\node at (0.2,0.2)[circle,fill,inner sep=1.5pt]{};

\draw (-3,0) node{$\psi$}; 
\draw (-1.4,-0.6) node{$0$}; 
\draw (-2.6,1.2) node{$\tau_2$};
\draw (0.2,0.6) node{$t_1$};
\end{tikzpicture}\ =\int^{\pi\Lambda}_{-\pi\Lambda}\frac{dk}{2\pi}\frac{e^{-i\epsilon_k(t_1+i\tau_2)}}{1+e^{\beta\epsilon_k}}= g(t_1-i(\beta-\tau_2)).
\end{equation}
\begin{equation}
\begin{tikzpicture}[thick,scale = 0.8,baseline={([yshift=-3pt]current bounding box.center)}]
\draw (-0.3+1,-0.2) arc (-90:90:0.1 and 0.2);

\draw (-0.3+1,0.2) --(-0.3-0.1-1.2,0.2) ;
\draw  (-0.3+1,-0.2) --(-0.3-0.1-1.2,-0.2) ;

\draw[thick,->,>=stealth]  (-0.3-0.1-1.2,0.2) --(-0.3-0.15,0.2) ;
\draw[thick,->,>=stealth]  (-0.3+1,-0.2) --(-0.3-0.15,-0.2) ;

\node at (-1,0.2)[circle,fill,inner sep=1.5pt]{};
\node at (0.2,0.2)[circle,fill,inner sep=1.5pt]{};
\draw  (-0.3-0.1-1.2,0.16)  arc (11:360-11:1 and 1);
\draw (-3,0) node{$\psi$}; 
\draw (-1,0.6) node{$t_2$}; 
\draw (0.2,0.6) node{$t_1$};
\end{tikzpicture}\ =\int^{\pi\Lambda}_{-\pi\Lambda}\frac{dk}{2\pi}\frac{e^{-i\epsilon_k(t_1-t_2)}}{1+e^{-\beta\epsilon_k}}= g(t_1-t_2).\ \ \ \ \ \ \ \ \ \ 
\end{equation}
We are mainly interested in the low-energy modes. Consequently, we further make the approximation by using a linear dispersion $\epsilon_k\approx k$ with a cutoff of the order of $\Lambda$. The integral can then be carried out explicitly which gives 
\begin{align}
g(t)\approx 2\int^\Lambda_\Lambda \frac{dk}{2\pi} \frac{e^{-ikt}}{1+e^{-\beta k}}=\frac{-e^{-\frac{\pi  t}{\beta }} B_{-e^{\Lambda \beta }}\left(1-\frac{i t}{\beta
   },0\right)+e^{\frac{\pi  t}{\beta }} B_{-e^{\Lambda \beta }}\left(\frac{i t}{\beta
   },0\right)+i \pi  \text{csch}\left(\frac{\pi  t}{\beta }\right)}{\pi  \beta }.\label{Gref}
\end{align}
Here $B_z(a,b)$ is the incomplete beta function defined as
\begin{align}
B_z(a,b)\equiv \int_0^z dx u^{a-1}(1-u)^{b-1}du, \notag
\end{align}
and the factor of $2$ comes from the summation over the left-moving and right-moving modes. At low-temperature limit $\Lambda \beta \rightarrow \infty$, using
 \begin{align}
     \text{lim}_{z\rightarrow-\infty}B_{z}(a,b)=\frac{(-1)^{a+1} \Gamma (a) \Gamma (-a-b+1)}{\Gamma (1-b)}
 \end{align}
and \eqref{Gref}, we recover the conformal Green's function  
 \begin{align}
    g(-i(\tau_1-\tau_2))=\frac{1}{\pi}\left(\frac{\pi}{\beta \sin \frac{\pi  (\tau_1-\tau_2) }{\beta }}\right),
 \end{align}
 as expected.

\section{Perturbative analysis in the short time limit}\label{B}
In this appendix, we give details for the calculation of the short-time action \eqref{action short}.

Defining $\theta_i=\frac{2\pi}{\beta}\tau_i$ and using the explicit formula \eqref{Gchi} \eqref{Gpsi} for the Green's function, we find
        \begin{align}
         \frac{(n-1)S^{(n)}}{N}=\frac{n V^2}{2^{3/2}}\frac{1}{(4\pi J^2 )^{1/4}} \frac{1}{\pi}\sqrt{\frac{\beta}{2\pi}}\left( \int_{\epsilon_\theta}^{\theta_0-\epsilon_\theta}d\theta_1\int_{\theta_0+\epsilon_\theta}^{2\pi-\epsilon_\theta}d\theta_2\  \left( \frac{1}{ \sin \frac{(\tau_1 - \tau_2)}{2} } \right)^{\frac{3}{2}}\right),
        \end{align}
with $\epsilon_\theta=2\pi\epsilon/\beta$. The integral can be carried out explicitly. After continuation to real time by $\tau_0 = \frac{\beta}{2} - 2it$ and taking $\epsilon \rightarrow 0$, we get
    \begin{equation}
    \begin{aligned}
\frac{S^{(2)}}{N}=\frac{V^2 }{2^{\frac{3}{2}} \pi ^{\frac{7}{4}}}&\sqrt{\frac{\beta }{J}}\left[
\frac{32 i \pi  t E\left(\left.\frac{i \pi  t}{\beta }\right|2\right)}{\beta }+\frac{32}{15} \cosh ^{\frac{5}{2}}\left(\frac{2 \pi  t}{\beta }\right) \,
   _3F_2\left(1,\frac{5}{4},\frac{5}{4};\frac{7}{4},\frac{9}{4};\cosh ^2\left(\frac{2
   \pi  t}{\beta }\right)\right)\right.\\ &\left. -\frac{16 \sqrt{\cosh \left(\frac{2 \pi  t}{\beta }\right)} \left(\pi  t \sinh
   \left(\frac{4 \pi  t}{\beta }\right) \, _2F_1\left(1,\frac{5}{4};\frac{7}{4};\cosh
   ^2\left(\frac{2 \pi  t}{\beta }\right)\right)-6 \beta \right)}{3 \beta }\right] +\text{const.}
\end{aligned}\label{resshort}
    \end{equation}
Here $E\left(\left.\phi\right|m\right)$ is the elliptic integral of the second kind and $_pF_q(a_1,...a_p;b_1,...b_q;z)$ is the generalized hypergeometric function. Since we are mainly interested in the time dependence, we do not give the explicit formula of the constant term. Taking the leading order contribution with $t\gg \beta$, we arrive the result quoted in the main text:
\begin{align}
\frac{S^{(2)}}{N}=\frac{8 V^2 \Gamma \left(\frac{3}{4}\right)^2 }{\pi ^{5/4}}\sqrt{\frac{\beta }{J}}\frac{t}{\beta}  + \textrm{const}.
\end{align}

\section{Factorization of the two-point function of twist operators}\label{C}

In this appendix we give more detailed argument for the factorization of the correlation function of the twist operators. In numerics, we observed that the backreactions of the twist operators are local. This means that we can approximate the Green's function of $\chi$ by 
\begin{equation}
    G_{T_LT_R} \approx G_{I} + \delta G_{ T_L} + \delta G_{T_R},
\end{equation}
where $G_{I}$ is the solution with no twist operator inserted (fig. \ref{long}(b)), where $I$ stands for identity, while $\delta G_{T_L}$ and $\delta G_{T_R}$ corresponds to the backreaction of the twist operator $T_L$ and $T_R$. The support of $\delta G_{T_L}$ and $\delta G_{T_R}$ are separated by a real time evolution $t$ much greater than $\beta$ on the contour. When we evaluate the action, this leads to
\begin{align}
    \frac{I_{\mathcal{C}}(T_LT_R)}{N}&= \frac{1}{2} \log \det G_{T_LT_R} - \frac{1}{2} \log \det \left(G_{0,\psi}^{-1}\right) 
    +   \int_\mathcal{C}ds\,ds'\  \frac{3J^2G_{T_LT_R}^4}{8}\notag\\
    &=\frac{1}{2} \log \det G_{I}+\frac{1}{2} \log \det (I+G^{-1}_{I}\circ \delta G_{ T_L})+\frac{1}{2} \log \det (I+G^{-1}_{I}\circ \delta G_{ T_R}) \notag\\&\ \ \ - \frac{1}{2} \log \det \left(G_{0,\psi}^{-1}\right) 
    +   \int_\mathcal{C}ds\,ds'\  \frac{3J^2(G_{I}+\delta G_{T_L})^4}{8}
    +   \int_\mathcal{C}ds\,ds'\  \frac{3J^2(G_{I}+\delta G_{T_R})^4}{8}\notag \\ &\ \ \ -  \int_\mathcal{C}ds\,ds'\  \frac{3J^2G_{I}^4}{8}+\frac{\delta I_\mathcal{C}}{N}\notag\\
    &=\frac{I_{\mathcal{C}}(T_L)}{N}+\frac{I_{\mathcal{C}}(T_R)}{N}-\frac{I_{\mathcal{C}}(I)}{N}+\frac{\delta I_\mathcal{C}}{N}.
\end{align}
Here we have separated out crossing terms of $\delta G(T_L)$ and $\delta G(T_R)$ into $\delta I_\mathcal{C}$. There are two kinds of diagrams in $\delta I_\mathcal{C}$. Terms from $G^4$ is of the form:
\begin{align}
    \int_\mathcal{C}dsds'\delta G_{T_L}^a\delta G_{T_R}^bG_I^{4-a-b}\sim O(e^{-t/\beta}),
\end{align}
Here we have $a>0$, $b>0$ and $a+b\leq 4$. Terms from the $\log \det$ term is also suppressed by $e^{-t/\beta}$. As an example:
\begin{align}
    \text{tr}\left[ \delta G_{T_L}\circ G^{-1}_I \circ \delta G_{T_R}\circ G^{-1}_I\right]\sim O(e^{-t/\beta}).
\end{align}
Since $G^{-1}(s,s')$ decays exponentially for large real-time separation. Combining these results, we have 
\begin{align}
    \frac{I_{\mathcal{C}}(T_LT_R)}{N}\approx\frac{I_{\mathcal{C}}(T_L)}{N}+\frac{I_{\mathcal{C}}(T_R)}{N}-\frac{I_{\mathcal{C}}(I)}{N}.
\end{align}
Realizing that $\exp(-I_{\mathcal{C}}(I))=Z^2$ is just the partition function, we find the saturation value of the entropy is just twice of the thermal R\'{e}nyi entropy for subsystem $\chi$, which means the factorization of twist operators.

\section{Entropy dynamics with an SYK bath}
In this appendix, we present results for a related model by replacing the Majorana chain bath by an large SYK bath. This is an extension of the results in \cite{penington2019replica}, where the authors studied the case with equal number of modes. We find similar results as the chain bath case. We now consider two SYK systems $\chi$ and $\psi$ with different number of modes $N_\chi$ and $N_\psi$. The Hamiltonian of the system is written as:
        \begin{align}\label{model}
        H=H_{\chi}+H_{\psi}+H_{\text{int}}= \sum_{i,j,k,l}\frac{J^\chi_{ijkl}}{4!}\chi_i\chi_j\chi_k\chi_l+\sum_{i,j,k,l}\frac{J^\psi_{ijkl}}{4!}\psi_i\psi_j\psi_k\psi_l+H_{\text{int}},
        \end{align}
        with the variances for $J^\chi_{ijkl}$ and $J^\psi_{ijkl}$ being:
        \begin{align}
        \overline{(J^\chi_{ijkl})^2}=\frac{3!J^2}{N_\chi^3},\ \ \ \ \ \ \  \overline{(J^\psi_{ijkl})^2}=\frac{3!J^2}{N_\psi^3}.
        \end{align}
        We add an interaction term $H_{\text{int}}$ that couples $\chi$ and $\psi$ system. In this model, we consider two types of interaction. One is of the "$\chi^2\psi^2$" form:
        \begin{align}
        H^{\chi^2\psi^2}_{\text{int}}=\sum_{i,j,k,l}\frac{V_{ijkl}}{4}\chi_i\chi_j\psi_k\psi_l,\ \ \ \ \ \ \overline{(V^\chi_{ijkl})^2}=\frac{2V^2}{N_\chi N_\psi^2},
        \end{align}
        and another is of the "$\chi\psi^3$" form:
        \begin{align}
            H^{\chi\psi^3}_{\text{int}}=\sum_{i,j,k,l}\frac{V_{ijkl}}{3!}\chi_i\psi_j\psi_k\psi_l,\ \ \ \ \ \ \overline{(V^\chi_{ijkl})^2}=\frac{3!V^2}{N_\chi N_\psi^2}.
        \end{align}
We define the ratio of fermion number to be $r \equiv N_\psi/N_\chi$. 

We again compute the second R\'{e}nyi entopy \eqref{Renyi} after the evolution of a TFD state. For the $\chi^2\psi^2$ case, the path-integral formalism now reads
\begin{align}
           e^{-S^{(2)}_{\chi_L,\chi_R}}&= \frac{1}{Z^2}\int \mathcal{D}\tilde{\Sigma}_\chi \mathcal{D}\tilde{\Sigma}_\psi \mathcal{D}\tilde{G}_\chi \mathcal{D}\tilde{G}_\psi\exp(-S_\mathcal{C}[\tilde{\Sigma},\tilde{G}])
       \end{align}
\begin{equation}
\begin{aligned}
           S_\mathcal{C}[\tilde{\Sigma},\tilde{G}]&= - \frac{N_{\chi}}{2} \log \det \left(G_{0,\chi}^{-1}-\tilde{\Sigma}_\chi\right) - \frac{N_{\psi}}{2} \log \det \left(G_{0,\psi}^{-1}-\tilde{\Sigma}_\psi\right)
          \\& +
           \int_\mathcal{C}ds\,ds'\  \left[\frac{N_\chi}{2} \left( \tilde{G}_\chi \tilde{\Sigma}_\chi-\frac{J^2\tilde{G}^4_\chi}{4}F \right)+\frac{N_\psi}{2}\left( \tilde{G}_\psi \tilde{\Sigma}_\psi-\frac{J^2\tilde{G}^4_\psi}{4}F \right) \right]\\& +
           \int_\mathcal{C}ds\,ds'\  \left[-\frac{N_\chi V^2}{4} \tilde{G}_\psi^2 \tilde{G}_\chi^2F \right],
\end{aligned}
\end{equation}
which gives the saddle point equation 
    \begin{align}\label{psi2chi2}
           G_\chi^{-1}&=G_{0,\chi}^{-1}-\Sigma_\chi,\ \ \ \Sigma_\chi=(J^2G^3_\chi+V^2G^2_\psi G_\chi)F,\\
           G_\psi^{-1}&=G_{0,\psi}^{-1}-\Sigma_\psi,\ \ \ \Sigma_\psi=(J^2G^3_\psi+V^2G^2_\chi G_\psi/r)F.
       \end{align}
Similar effective action and saddle point equation can be worked out straightforwardly for the $\chi\psi^3$ case.

The $r=1$ case with $\chi^2\psi^2$ interaction has been studied in \cite{gu2017spread,penington2019replica}, where no transition is found in the canonical ensemble. The transition appears if we instead consider the micro-canonical ensemble \cite{penington2019replica}. 

Here we instead focus on large $r$ limit, whose equilibrium physics and quench dynamics have been studied in \cite{chen2017tunable,zhang2019evaporation,almheiri2019universal}. Since the qualitative features (the short time linear growth and tirst order transition regardless of the strength $V^2/J^2$. In contrary, for the $\chi\psi^3$ interaction, similar to the chain case, there is no consistent exact replica diagonal solution and the transition only appears for small $V^2/J^2$. 

 \end{appendices}

\bibliographystyle{JHEP}

\bibliography{cite}

\end{document}